\documentclass[10pt,preprint]{aastex}
\newcommand{\Bw}{$B_{W}$}
\newcommand{\Ks}{$K_{S}$}
\newcommand{\chisq}{$\chi^{2}$}
\newcommand{\chisqred}{$\chi^{2}_{\nu}$}

\newcommand{\deltachisq}{$\Delta \chi^{2}$}
\newcommand{\Bootes}{Bo\"otes}
\newcommand{\Teff}{T$_{\rm eff}$}
\usepackage{graphicx}
\usepackage{url}
\usepackage{lscape}
\usepackage{mathtools} 
\usepackage{subfigure}  
\usepackage{rotating}
\usepackage{natbib}
\usepackage{lscape}
\DeclarePairedDelimiter{\abs}{\lvert}{\rvert}

\begin{document}


\begin{abstract} 

We classify the spectral energy distributions (SEDs) of 431,038 sources in the 9 sq. deg \Bootes\ field of the NOAO Deep Wide-Field Survey (NDWFS).  There are up to 17 bands of data available per source, including ultraviolet (\emph{GALEX}), optical (NDWFS), near-IR (NEWFIRM), and mid-infrared (IRAC/MIPS) data, as well as spectroscopic redshifts for $\sim$20,000 objects, primarily from the AGN and Galaxy Evolution Survey (AGES).  We fit galaxy, AGN, stellar, and brown dwarf templates to the observed SEDs, which yield spectral classes for the Galactic sources and photometric redshifts and galaxy/AGN luminosities for the extragalactic sources.  The photometric redshift precision of the galaxy and AGN samples are  $\sigma/(1+z)=0.040$ and $\sigma/(1+z)=0.169$, respectively, with the worst 5\% outliers excluded.  Based on the \chisqred\ of the SED fit for each SED model, we are able to distinguish between Galactic and extragalactic sources for sources brighter than $I=23.5$.   We compare the SED fits for a galaxy-only model and a galaxy+AGN model.  Using known X-ray and spectroscopic AGN samples, we confirm that SED fitting can be successfully used as a method to identify large populations of AGN, including spatially resolved AGN with significant contributions from the host galaxy and objects with the emission line ratios of ``composite'' spectra.  We also use our results to compare to the  X-ray, mid-IR, optical color and emission line ratio selection techniques.   For an $F$-ratio threshold of $F>10$ we find 16,266 AGN candidates brighter than $I=23.5$ and a surface density of $\sim1900$ AGN deg$^{-2}$. 

\end{abstract}


\title{A UV to Mid-IR Study of AGN Selection}
\author{
{Sun Mi Chung}\altaffilmark{1},
{Christopher S. Kochanek}\altaffilmark{1},
{Roberto Assef}\altaffilmark{2},
{Michael J. Brown}\altaffilmark{3},
{Daniel Stern}\altaffilmark{4},
{Buell T. Jannuzi}\altaffilmark{5}, 
{Anthony H. Gonzalez}\altaffilmark{6},
{Ryan C. Hickox}\altaffilmark{7},
{John Moustakas}\altaffilmark{8}
}
\altaffiltext{1}{Department of Astronomy, The Ohio State University, 140 W.\ 18th Ave., Columbus, OH 43210, USA}
\altaffiltext{2}{N\'ucleo de Astronom\'ia de la Facultad de Ingenier\'ia, Universidad Diego Portales, Av. Ej\'ercito Libertador  441, Santiago, Chile} 
\altaffiltext{3}{School of Physics, Monash University, Clayton, Victoria 3800, Australia}
\altaffiltext{4}{Jet Propulsion Laboratory, California Institute of Technology, 4800 Oak Grove Drive, Mail Stop 169-221, Pasadena, CA 91109, USA}
\altaffiltext{5}{Department of Astronomy and Steward Observatory, University of Arizona, Tucson, AZ 85721, USA}
\altaffiltext{6}{Department of Astronomy, University of Florida, Gainesville, FL 32611, USA}
\altaffiltext{7}{Department of Physics and Astronomy, Dartmouth College, 6127 Wilder Laboratory, Hanover, NH 03755, USA}
\altaffiltext{8}{Department of Physics and Astronomy, Siena College, 515 Loudon Road, Loudonville, NY 12211, USA}

\section{Introduction} 

\label{sec:Intro}
Active Galactic Nuclei (AGN) present a wide variety of observational properties, which can be exploited to identify them in large surveys using a range of selection techniques.  However, most selection methods are based on the characteristics in a limited wavelength region.  This leads to selection methods that are  incomplete for some AGN luminosities, classes, or redshifts.  For example, the early single (optical) color searches have evolved into multi-color optical surveys such as the Sloan Digital Sky Survey (SDSS), which has successfully identified $\sim10^{5}$ quasars based largely on their optical colors \citep{Richards2002, Richards2005, Schneider2007, Schneider2010}.  The color selection methods of \citet{Richards2002} achieved a completeness of $\sim$90\% for unobscured, unresolved AGN with $i<19$, $z<5$, and emission dominated by the AGN, with an efficiency (number of quasars/number of candidates) of $\sim$65\%.  However, using $ugriz$ colors also presents difficulties in distinguishing between F stars and quasars at $z\sim2$.  Since quasar activity is known to peak near this redshift \citep{Richards2006, Assef2011}, this presents a problem for studying the accretion history of the universe.   Using broadband optical colors to find AGN also relies on the roughly universal UV/optical power-law shape of direct emission from an accretion disk to distinguish AGN from the primarily photospheric emission of stellar populations in this wavelength range.  Obscured, type-2 AGN are more difficult to identify using optical colors, due both to the distortion of the shape of the AGN continuum and the increasingly significant host galaxy contribution.  Similarly, low accretion rate AGN are dominated by their host emission \citep[e.g.][]{Hopkins2009}.

X-ray emission is sensitive to both unobscured and moderately obscured AGN, with the added advantage that X-ray emission almost always implies the presence of AGN except at low luminosities ($L_{X}\lesssim10^{40}$ erg s$^{-1}$) where binary contributions become important \citep{Shapley2001}.  However, \emph{Chandra} and \emph{XMM-Newton} surveys are biased against highly obscured AGN whose soft X-rays are absorbed by large columns of gas and dust \citep[e.g.,][]{Polletta2006}.  Hard X-ray observations largely avoid this problem, but until the recent launch of \emph{NuSTAR} \citep{Harrison2013,Alexander2013}, with its greater sensitivity to hard X-rays, there were only shallow surveys for the highly obscured, Compton-thick AGN, such as the \emph{Swift} BAT Survey \citep{Tueller2008}.  Radio surveys can also be used to search for AGN; however, typical radio surveys are limited to a very small population of AGN, since only $\sim$10\%-20\% are radio-loud \cite[e.g.][]{Kellermann1989, Urry1995, Stern2000, Ivezic2002}.

Mid-infrared (mid-IR) surveys are sensitive to both type-1 and type-2 AGN.  The accretion disk power law spectrum extends into the rest-frame near-IR and is less affected by extinction.  Additionally, at rest-frame mid-IR and longer wavelengths, there is emission by hot dust in the ``torus'' where the UV and optical emission from the accretion disk is absorbed and reprocessed \citep[e.g.,][]{Sanders1989,Netzer2007}.  These contributions roughly produce a power-law spectrum in the mid-IR that is  easily distinguished from the infrared spectra of stars or relatively normal galaxies which are essentially falling Rayleigh-Jeans spectra regardless of stellar population or age at these wavelengths.  At lower redshifts ($z\lesssim1.5$) mid-IR selection triggers on dust emission, while at higher redshifts ($z\gtrsim1.5$) it really triggers on the direct emission from the disk.  As a result, many AGN occupy a region of mid-IR color space that is well separated from that of galaxies and stars \cite[e.g.,][]{Lacy2004, Stern2005, Alonso2006, Donley2007}, which was initially demonstrated using data from the four bands (3.6\micron\ to 8.0\micron) of \emph{Spitzer}/IRAC.  The \emph{Wide-Field Infrared Survey Explorer} \cite[\emph{WISE};][]{Wright2010} can be used to extend these methods over the full sky \citep{Mateos2012, Stern2012, Assef2013, Yan2013}.  However,  mid-IR color selection methods fail when the host galaxy contribution becomes large, or a strong H$\alpha$ line lies in the 3.6\micron\ band at $z\sim4$ and distorts the colors \citep{Assef2010}.  For quasars that are as faint as W2$\sim$15.5 in the WISE 4.6\micron\ band, galaxy contamination becomes significant and mid-IR color selection becomes more difficult \citep{Assef2013}.  

Each technique for identifying AGN has its own set of selection effects and it is important to understand the differences between AGN samples selected with the various methods.  For example, AGN that are detected using X-ray, radio, or mid-IR colors have different characteristic Eddington ratios, clustering properties, and host galaxy morphologies and host galaxy masses. \citet{Hickox2009}  studied a sample of $\sim$600 AGN at $0.25<z<0.8$ with spectroscopic redshifts from the AGES and Galaxy Evolution Survey \cite[AGES;][]{Kochanek2012}, and radio, X-ray, and mid-IR data from the Westerbork Synthesis Radio Telescope, the \emph{Chandra} X\Bootes\ Survey, and the \emph{Spitzer} IRAC Shallow Survey, respectively.  With modest overlap between the radio, X-ray, and mid-IR AGN samples, \citet{Hickox2009} found that radio AGN tend to be optically red and have massive host galaxies (i.e. red sequence galaxies), while X-ray AGN span a large range of host masses and colors but show a peak in the ``green valley'', and mid-IR AGN tend to be slightly optically bluer than the X-ray AGN.  \citet{Griffith2010} arrived at a similar conclusion by examining the host galaxy morphologies of radio, mid-IR, and X-ray selected AGN, and found that radio-selected AGN tend to have early-type hosts, while mid-IR and X-ray selected AGN are more likely to have disky host galaxies.  This may represent a sequence in evolution of the quasar phase, which is thought to be triggered by a burst of star formation that funnels gas into the central engine, which ultimately quenches the star formation, transforming the optical colors from blue to red \citep{Hopkins2007, Cardamone2010, Povic2012}.

Traditionally, extragalactic surveys have either been wide and shallow, or narrow and deep.  Wide/shallow surveys such as the 2 Micron All-Sky Survey \cite[2MASS;][]{Skrutskie2006}, the Sloan Digital Sky Survey \cite[SDSS;][]{Abazajian2009}, and the 2 Degree Field Galaxy Redshift Survey \cite[2dF;][]{Colless2001} have allowed studies of large numbers of nearby galaxies and contributed significant discoveries about galaxy properties in the low redshift universe and luminous AGN at all redshifts \citep[e.g.][]{Richards2005,Huchra2012}.  Examples of deep/narrow surveys include the Great Observatories Origins Deep Survey \cite[GOODS;][]{Dickinson2003} and the Cosmological Evolution Survey \cite[COSMOS;][]{Scoville2007}.  GOODS covers $\sim$300 arcmin$^{2}$ with deep observations from \emph{Spitzer}, \emph{Hubble}, \emph{Chandra}, \emph{Herschel} and \emph{XMM-Newton}, as well as ground-based facilities.  COSMOS is a wider survey, covering $\sim$2  deg$^{2}$ with imaging from both space-based and ground-based telescopes, including \emph{Hubble}, \emph{Spitzer}, \emph{GALEX}, \emph{Chandra}, \emph{NuSTAR}, Subaru, VLT, and others.   The relative rarity of AGN makes it difficult however to fully characterize AGN in deep fields or to fully account for faint AGN in shallow fields. 

While these (and other) extragalactic surveys have provided a tremendous contribution to our knowledge of both nearby and distant galaxy populations, there have been few surveys on $\sim$10 deg$^{2}$  scales.  The NOAO Deep Wide-Field Survey \citep[NDWFS;][]{Jannuzi1999} in the 9 deg$^{2}$ \Bootes\ field allows us to explore a cosmologically significant volume with data that is both deep and extensive in wavelength coverage.  The large survey volume means that studies in the \Bootes\ field are less affected by cosmic variance or small number statistics for rare, faint objects that would not be found in either deep/narrow or wide/shallow surveys.  In addition to near-IR and optical data from NDWFS, there is a wealth of ancillary data in the \Bootes\ field.  This includes additional optical ground-based data \citep{Cool2007,Bian2013}, X-ray data from the \emph{Chandra X-ray Observatory}  \citep{Murray2005, Kenter2005, Brand2006}, UV data from the \emph{Galaxy Evolution Explorer} \cite[\emph{GALEX};][]{Hoopes2004}, and mid-IR data from the \emph{Spitzer Space Telescope} \citep{Eisenhardt2004, Ashby2009, Jannuzi2010}.   There is also spectroscopic data from the AGN and Galaxy Evolution Survey \cite[AGES;][]{Kochanek2012}, which provides redshifts for $\sim$18,000 galaxies ($I<20$) and $\sim$4,700 AGN candidates ($I<22.5$).

In this paper, we examine and classify the UV to mid-IR spectral energy distributions (SEDs) of 431,038 sources in the \Bootes\ survey region.  By fitting galaxy, AGN, and stellar templates to the observed SEDs, we are able to isolate AGN candidates from galaxy and stellar populations using a much broader wavelength range than the common AGN selection methods discussed previously.  In \S\ref{sec:data} we outline the data that are used in this work.  In \S\ref{sec:analysis-SEDfitting} and \S\ref{sec:analysis-droppedfilters} we compare photometric to spectroscopic redshifts and discuss in detail the SED fitting procedure and how our final sample of SED fits were assembled.  In \S\ref{sec:analysis-galorexgal} we illustrate the separation of Galactic and extragalactic sources based on \chisqred\ statistics and we investigate the difficulty of classifying sources at faint magnitudes.   Section \S\ref{sec:AGNcandidates} presents how AGN candidates may be selected based on the results of their SED fits as galaxy and galaxy+AGN models and the sensitivity of this SED selection to the relative strength of the underlying host galaxy. In \S\ref{sec:compare}, we compare the SED selection of AGN candidates to mid-IR and optical color selections. In \S\ref{sec:surface_density} we compare the surface densities of different AGN samples. Finally in \S\ref{sec:future} we summarize our most important results and comment on future work.   Throughout this work we assume a flat $\Lambda$CDM cosmology with $H_{0}=70$ km s$^{-1}$ Mpc$^{-1}$, $\Omega_{M}=0.28$, and $\Omega_{\Lambda}=0.72$.

\section{Data}  
\label{sec:data} 

The data used in this work consist of the extensive multi-wavelength imaging of the 9 deg$^{2}$ NDWFS \Bootes\ field and optical spectroscopy from AGES \citep{Kochanek2012} and Hickox et al. (private communication).   The following sections provide brief descriptions of the various datasets used in this paper.  Data from  NDWFS  (\Bw, $R$, $I$, $K$) and NEWFIRM ($J$, $H$, $K_{s}$) are based on the Vega magnitude system, and data from \emph{GALEX},  $U_{S}$, $Y$, and $z$ bands are based on the AB system.  All magnitudes reported throughout  this paper are kept in their native zeropoint systems and refer to $6^{\prime\prime}$ diameter aperture magnitudes unless otherwise stated.  The 3$\sigma$ magnitude limits quoted for each of the filters is based on calculating the signal-to-noise (S/N) ratio from the fractional flux error.  The median S/N is computed in bins of 0.25 mag, and a simple linear interpolation is used to obtain the magnitude where S/N=3$\sigma$.  

Objects were detected using SExtractor 2.3.3 \citep{Bertin1996} run on the $I$-band images from the NDWFS third data release. To obtain consistent aperture photometry across the full \Bootes\ field, the \Bw, $R$, $I$, $Y$, $H$ and \Ks\ images were smoothed to a common PSF of 1\farcs35 FWHM while the $U_{S}$, $z$ and $J$ images were smoothed to a common PSF of 1\farcs60 FWHM. We measured aperture photometry for each object using our own code. SExtractor segmentation maps were used to exclude flux associated with neighboring objects, and we corrected the photometry for missing pixels using the mean flux per pixel measured in a series of annuli surrounding each object. Uncertainties were determined by measuring photometry at $\simeq$100 positions within $2^{\prime}$ of the object position and finding the range that encompassed 68\% of the measurements.  Saturation mainly occurs for sources brighter than $R\simeq17$ in typical NDWFS exposures \citep{Jannuzi1999}, which corresponds roughly to $I\simeq16.5$.  For a more extensive description of the photometric catalogues we refer the reader to \citet{Brown2007}.          
           
\subsection{X-ray Observations}

X\Bootes\  is an X-ray survey of the 9 deg$^{2}$  NDWFS \Bootes\ field, with data taken from ACIS-I onboard the \emph{Chandra X-ray Observatory} \citep{Murray2005, Kenter2005, Brand2006}.  The integration time was 5 ks per position and the survey identified 3293 point sources with $\ge$4 counts.  This corresponds to a flux of $7.8\times10^{-15}$ ergs cm$^{-2}$ s$^{-1}$ in the 0.5-7 keV range assuming a standard unabsorbed AGN X-ray spectrum.  However, most 4-count sources actually correspond to more numerous sources modestly fainter than this flux limit brought into the catalog by Poisson fluctuations in their counts \cite[see][]{Kenter2005}.

\subsection{UV Imaging}

\emph{GALEX} observed the field in its far-UV (FUV) and near-UV (NUV) bands with 3$\sigma$ magnitude limits of 24.7 and 25.5 mag (AB), respectively.   

\subsection{Optical Imaging}

The optical data consists of the NDWFS \Bw, $R$, and $I$ bands \citep{Jannuzi1999}, the $z$ band \cite[z\Bootes;][]{Cool2007}, and the $U_{S}$ band from the Large Binocular Telescope \cite[LBT;][]{Bian2013}.  The 3$\sigma$ depths of the $U_{S}$, \Bw, $R$, $I$, and $z$ bands are 24.9 (AB), 25.2 (Vega), 23.9 (Vega), 22.9 (Vega), and 22.3 (AB) mag, respectively.  

\subsection{Near-Infrared}

Near-IR data from the Infrared \Bootes\ Imaging Survey \cite[IBIS;][]{Gonzalez2010} provide photometry in the $J$, $H$, and \Ks\ bands with 3$\sigma$ depths of 21.1, 20.1, 18.9 (Vega).  There is also $K$-band data from the original NDWFS survey, though this is significantly shallower than the NEWFIRM data, with a 3$\sigma$ limit of 17.4 mag (Vega).  Finally, the $Y$-band data from the LBT \citep{Bian2013} has a 3$\sigma$ limit of 22.2 mag (AB).

\subsection{Mid-Infrared} 

Mid-IR observations from the \emph{Spitzer} Deep, Wide-Field Survey \cite[SDWFS;][]{Ashby2009} and the MIPS AGN and Galaxy Evolution Survey \cite[MAGES;][]{Jannuzi2010} provide data in the four IRAC channels and the 24\micron\ MIPS channel.  SDWFS includes data from the earlier IRAC Shallow Survey \citep{Eisenhardt2010}.  The 3$\sigma$ depths of the four IRAC bands ([3.6], [4.5], [5.8], [8.0]) and the MIPS 24\micron\ band are 19.3, 18.7, 16.8, 16.1, and 11.8 mag, respectively.  

\subsection{Spectroscopic Data}

In addition to the photometric catalogs, we use spectroscopic data from AGES \citep{Kochanek2012}, which was carried out with  MMT/Hectospec \citep{Roll1998, Fabricant2005}.   The AGES galaxy samples were designed to sample the broad range of galaxy colors with well-defined samples at all wavelengths from \Bw\ to 24\micron.  It is complete for $I<18.5$ and sparsely sampled for $18.5<I<20$ galaxies, with redshifts obtained for $\gtrsim30$\% of all $I<20$ galaxies.  The AGES limiting magnitude for AGN was significantly fainter, reaching $I<22.5$ mag for point-like sources in the $I$-band.  The AGN targets were selected based on several different criteria, including point-like sources with MIPS emission, mid-IR colors, and X-ray or radio emission.  A total of $\sim$18,000 redshifts were collected for $I<20$ galaxies and $\sim$4,700 redshifts for $I<22.5$ AGN candidates.  We extend this spectroscopic sample with an additional $\sim$1000 Hectospec redshifts from Hickox et al. (private communication).  The magnitude of a field $L_{\star}$ galaxy at $z\sim1$ is $i^{\prime}\sim21.2$ \citep{Gabasch2006}, roughly corresponding to $I\sim20.8$ (Vega).  This is fainter than the limiting magnitude of the AGES and Hickox galaxy samples and we can consider essentially all $z\gtrsim1$ spectroscopic sources to be AGN.

We start with $\sim$two million sources that have either $I<23.5$ or $[4.5]<18$.  These two limits are broadly comparable for typical sources, but by using both we avoid biases against optically faint sources such as high redshift galaxies/AGN or brown dwarfs.   The optical limiting magnitude corresponds to a galaxy luminosity of 0.1$L^{*}$ at $z=1$ \cite[e.g.,][]{Gabasch2006}.   We next require that each source has data in at least ten different filters, with a minimum of five detections at S/N$>$3.  Since we fit SED models to the fluxes, negative flux estimates are simply included as measurements, which is more statistically correct than including them as one-sided upper bounds.  Photometric uncertainties that are too small compared to the systematic residuals typical of the SED fits will bias the fits.  Therefore we assign a minimum photometric error of 0.05 mag.  Because the NDWFS $I$-band subfields overlap, there can be up to 4 detections for individual objects. To select the best detection, we reject objects that are outside nominal subfield boundaries and then (if needed) select the detection with the highest S/N. There are regions within the fields that have poor data (e.g., bad pixels, saturated pixels) or increased backgrounds (e.g., bright objects halos). We excluded objects in these regions based on ``flagdeep'', which identifies objects with good faint object photometry.  These criteria left us with a sample of 431,038 sources which we fit with galaxy, AGN, and stellar templates.   Among the $\sim$1.5 million sources that we exclude from our sample, roughly 25\% are lost due to ``regional'' issues associated with ``flagdeep'', while the remaining sources are lost due to not having enough bands that meet the S/N criterion.



\section{Analysis}

This section discusses the SED fitting procedure and compares the resulting photometric redshifts with spectroscopic redshifts for $\sim$20,000 sources.  We present the \chisq\ statistic per degree of freedom (\chisqred) of the galaxy+AGN and stellar/brown dwarf SED models and demonstrate that examining \chisqred\ is an effective way to separate Galactic and extragalactic sources.  The results of the final fits are reported in Table~\ref{table:results}.

\subsection{SED fitting}
\label{sec:analysis-SEDfitting}

We use the empirically derived SED templates of \citet{Assef2010} to fit the sources.  The templates extend from 0.03 to 30\micron.  There are three galaxy templates, corresponding to ``elliptical'' (old stellar population), ``spiral'' (on-going star formation), and ``irregular'' galaxies (starburst population), and a single AGN template.  The AGN template is fit with a variable amount of internal reddening and both galaxy and AGN templates are fit with an IGM absorption model that is a fixed function of redshift \citep{Stengler1995,Fan2006}.  We do not fit the IGM absorption as a free parameter because this worsens the photometric redshifts \citep{Assef2010}.  In addition to the galaxy and AGN templates, we fit all sources with the stellar atmosphere models of \citet{Castelli2004} and as brown dwarfs with effective temperatures of 2500K, 2000K, 1500K, and 1000K from the \citet{Allard2007} models.  We linearly interpolate between stellar templates of adjacent temperature/spectral class to create a one parameter sequence in stellar temperature for both dwarf and giant spectral classes.

The templates are fit to the data using the publicly available code of \citet{Assef2010}, which fits non-negative linear combinations of the templates to the data.  An $R$-band galaxy luminosity function based on the Las Campanas Redshift Survey \citep{Lin1996} is used as a luminosity prior on the galaxy templates in order to avoid unlikely luminosities.  No such prior is applied to the luminosity of the AGN component in the SED.   The data are fit once with only the three galaxy templates, then again with the galaxy templates and the AGN component, and a final time with only the stellar and brown dwarf templates.  The redshift parameter of the fits is allowed to vary between 0 and 3 in steps of 0.01 when fitting for galaxies and AGN, and is fixed at $z=0$ for stars and brown dwarfs.

The goodness-of-fit is measured by \chisqred\, which is simply \chisq\ divided by the degrees of freedom (DOF or $\nu$). A good model fit should yield \chisqred $\simeq1\pm\sqrt{2/\nu}$.  Values significantly smaller than unity indicate that the data are being over-fit, while values much larger than unity signify a poor model fit.  For $N$ bands of data, including upper limits, the galaxy, galaxy+AGN, and stellar models have $\nu=N-4$, $N-6$, and $N-2$ degrees of freedom, respectively.  The galaxy model parameters are the three template luminosities and the redshift.  The galaxy+AGN  model adds the luminosity and extinction of the AGN component.  The stellar models depend on the total flux and the ``temperature'' parameter.  

We fit each object once using only the three galaxy templates, then a second time adding the AGN template.  We then determine whether the inclusion of the AGN template significantly improves the fit.  The \chisq\ value cannot be used to distinguish whether an SED is better fit with or without the AGN component because the two additional model parameters will always result in a ``better'' fit with a smaller \chisq.  In order to test whether a significant improvement has occurred, we calculate the $F$-ratio
\begin{equation} 
F = \frac{\chi^{2}_{\rm Gal}-\chi^{2}_{\rm Gal+AGN}}{\chi^{2}_{\rm Gal+AGN}}\cdot \frac{\nu_{\rm Gal+AGN}}{\nu_{\rm Gal}-\nu_{\rm Gal+AGN}},
\end{equation} 
which compares the change in \chisq\ to the change in degrees of freedom.  For Gaussian uncertainties, the probability distribution of $F$ is known and we can assign a probability to the improvement in the fit being significant.  For example, if $\nu_{\rm Gal}=11$ (implying that $\nu_{\rm Gal+AGN}=9$) and $F=10$, there is only a 0.5\% probability that the  \chisq\ improvement is a chance occurrence, and we have high confidence that the \chisq\ improvement is due to the presence of an AGN.  In practice, we must be more careful because the residuals of the fits are almost certainly not simple Gaussians.   

While the $F$-ratio and associated probability distribution (or the $F$-test) quantifies the probability that the SED of a particular object is better fit with an AGN component, we must also take into consideration the relative rarity of AGN compared to galaxies when constructing statistical samples.  Suppose $P(F)$ is the probability that the source is better fit as a galaxy rather than as a quasar by chance.  In a sample of $N_{G}$ galaxies and $N_{Q}$ quasars, we would expect to find $P(F) N_{G}$ galaxies that are falsely characterized as quasars and $(1-P(F)) N_{Q}$ quasars.  The number of sources above and below a given $F$ value can be estimated as $N(<F) = (1-P(F))N_{G} + P(F) N_{Q}$ and $N(>F) = P(F) N_{G} + (1-P(F))N_{Q}$, respectively. A crude estimate of the quasar fraction is then
\begin{equation}\label{eq:purity}
\frac{N_{Q}}{N(>F)} \simeq 1 - \frac{P}{(1-2P)}\left(\frac{N(<F)}{N(>F)}-1\right).
\end{equation}
For example, based only on the probability of a false positive associated with the $F$-ratio, one might naively expect that for $F(>3)$ (or $P=0.055$), there is an implied purity of $N_{Q}/(N(>F))\simeq95\%$.  However, the ratio of sources above and below $F=3$ is $N(>F)/N(<F)\simeq19\%$, and Equation~\eqref{eq:purity} would suggest an expected purity of $N_{Q}/N(>F)\simeq75\%$.   Even for $F>8$ (naively 99.9\% pure), Equation~\eqref{eq:purity} suggests a purity of 98\%.   This somewhat oversimplifies the problem, but it emphasizes the consequence of searching for relatively rare sources in large populations.

There are $\sim$20\% of sources in our sample that have $F<0$ because the data are slightly worse fit by the Galaxy+AGN model than by the galaxy-only model.  This occurs when the luminosity of the galaxy-only model is too high and there is a large penalty from the luminosity prior.   To solve this problem, the code includes an AGN component in order to decrease the galaxy luminosity, even at the expense of having a slightly worse \chisq\ for the fit to the SED because we optimize on the \chisq+prior, rather than just the \chisq.   Of the $F<0$ SED fits, more than 99\% have $\chi^{2}_{\rm Gal}-\chi^{2}_{\rm Gal+AGN} > -5$, which means the addition of an AGN component has not significantly improved the overall fit.  We interpret these sources as pure galaxies or galaxy-dominated SEDs with little to no AGN contribution.

\subsection{Anomalous Data Points}
\label{sec:analysis-droppedfilters}

\begin{figure}[!ht]
\centering
\subfigure{\includegraphics[width=3in]{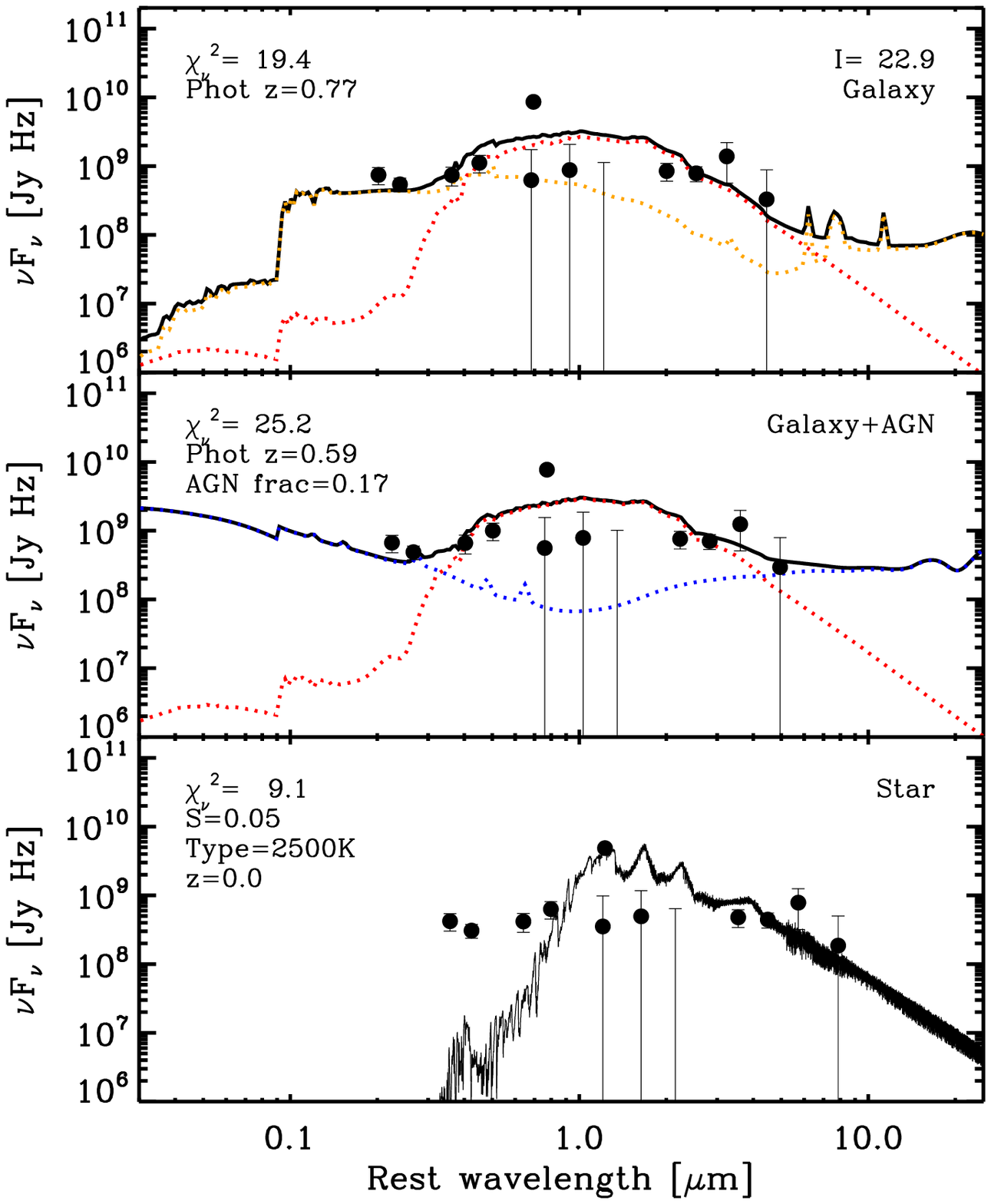}}
\subfigure{\includegraphics[width=3in]{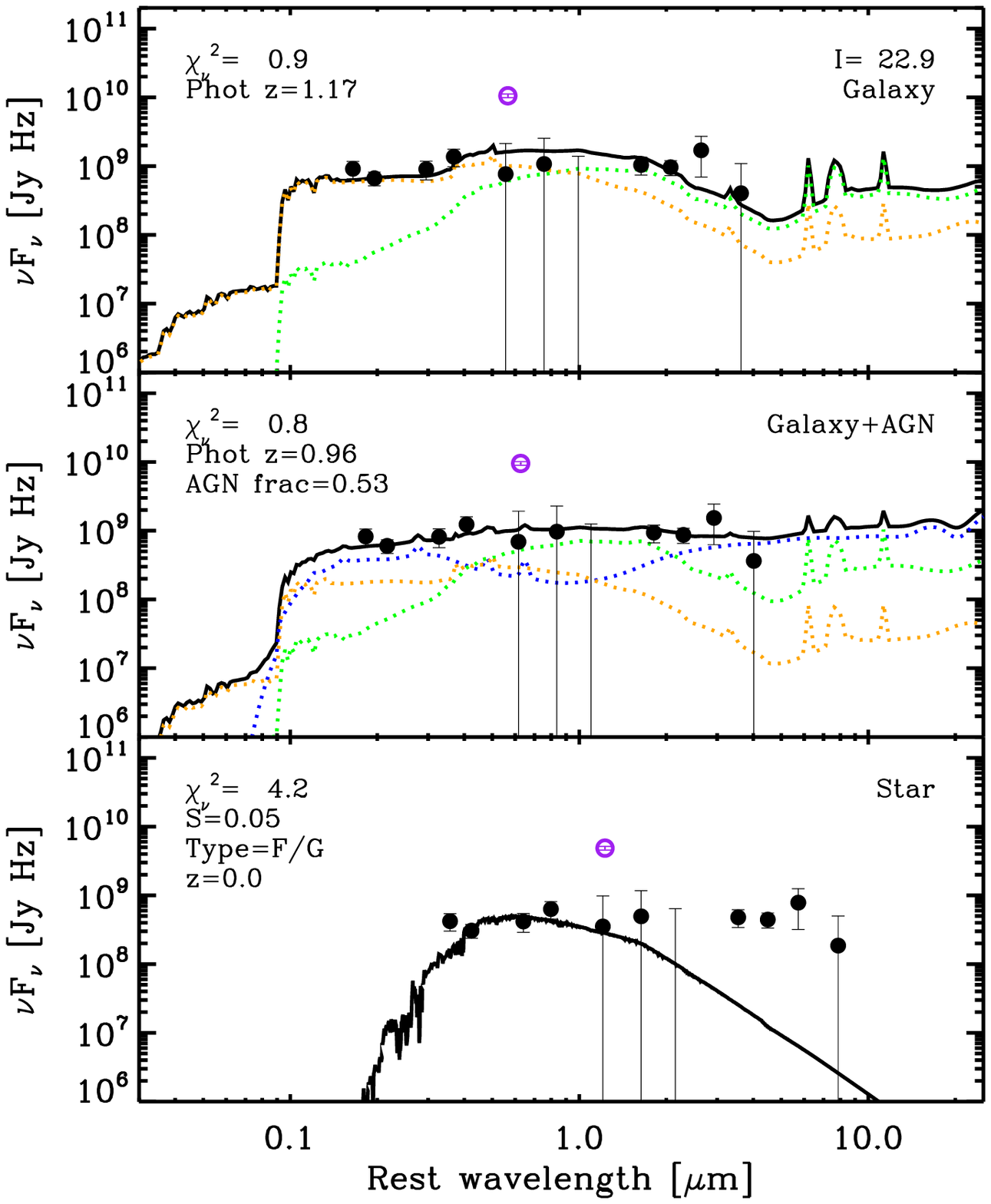}}
\caption{\scriptsize{Left: SED of a galaxy with an anomalous data point in the $Y$-band, with coordinates $\alpha$=216.40562 deg and $\delta$=32.32618 deg.  Due to the small photometric error of the $Y$-band measurement relative to the other bands, the SED fit does not characterize the true shape of the galaxy SED, leading to the large \chisqred\ in the left panels.  The top, middle, and bottom panels are fit with galaxy templates, galaxy+AGN templates, and a stellar or brown dwarf template, respectively.  The \chisqred\, photometric redshift, AGN fraction (ratio of AGN luminosity to total luminosity), stellarity index ($S$), and star or brown dwarf template type, are included.  The dotted blue, red, green, and orange lines show the AGN, elliptical, spiral, and irregular galaxy SED models, respectively.  Right: The $Y$-band measurement (purple open circle) is excluded from the fit, which results in a significantly improved fit, as illustrated by the large improvements in \chisqred. }}
\label{fig:SED_drop}
\end{figure}

One of the challenges of working with a large dataset is to confirm the fidelity of the SED fits for all sources.  With photometry for $\sim$430,000 sources from multiple telescopes, filters, and epochs, it is inevitable that some fraction of objects have badly fit SEDs due to inconsistent or bad photometry in one or more bands.  Problematic photometry can arise from either instrumental (e.g. stray light, cosmic rays), or astrophysical causes (e.g. variability, supernova).  Source confusion can also be an issue, especially for the \emph{GALEX} bands where the PSF is $\sim6^{\prime\prime}$.   One way to minimize the impact of outliers is to sequentially exclude data from a single filter and re-fit the SED to the remaining data.  If the results of the original fit are stable, then dropping any single band should not dramatically alter the SED fit, redshift, or the \chisq\ goodness-of-fit.  If, however, a single band is inconsistent with the data from all the other filters and is driving the results, then we can identify the problematic filter and use the results excluding the problematic filter.  Figure~\ref{fig:SED_drop} illustrates this with a galaxy that is badly fit by all templates due to an anomalous $Y$-band measurement. The SED fits in the left panel use all the bands, including the $Y$-filter, with the results for the galaxy templates, galaxy+AGN, and stellar templates shown from top to bottom.  In this case, dropping the $Y$-band data point greatly improves the fit (from \chisqred=19.4 to \chisqred=0.9) and the resulting SED fits are shown in the right panel of  Figure~\ref{fig:SED_drop}.   Data points with small error bars tend to be dropped more frequently because they cause large changes in \chisq.   While in the example shown in Figure~\ref{fig:SED_drop}, the SED model changes significantly due to the dropped filter, this is not always the case.  Sometimes a filter with a small error bar is dropped, thereby significantly improving the \chisq\ of the fit, yet the final SED model remains approximately the same. 

\begin{figure*}[!t]
\centering
\includegraphics[scale=0.6]{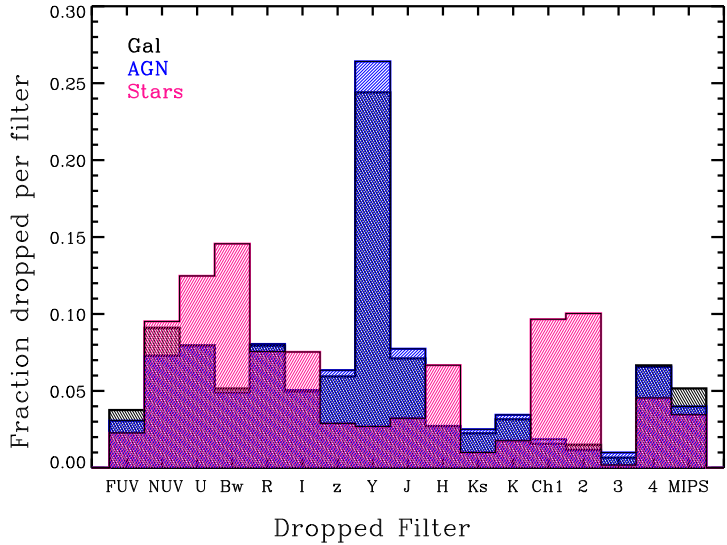}
\caption{\scriptsize{Distribution of dropped filters for the galaxy, AGN, and stellar populations (black, blue, and dark pink histograms, respectively).  Note that this shows the distribution of dropped filters among the 18\% of sources in our sample that have a single filter excluded from their final SED fits.}} 
\label{fig:dropped_filter}
\end{figure*}

Figure~\ref{fig:dropped_filter} shows the distribution of the dropped filters with band pass that produce a greater than $3\sigma$ improvement in the \chisq\ of the SED fits, where $\sigma=\sqrt{2\nu}$. We show the results for the galaxy, AGN, and stellar samples. The histogram for each sample is normalized to the total number of sources that are dropped.  The stellar sample includes sources that are better fit by stellar SED models rather than galaxy or AGN models.  The remaining extragalactic sources are separated into ``galaxy'' and ``AGN'' samples, where the AGN sample is comprised of sources with SED fits that show an improvement of \deltachisq $>20$\ after an AGN component is introduced to the fit.  While this is neither a pure nor a complete AGN sample, it is sufficient for the purpose of comparing the distribution of dropped filters among galaxy and AGN-dominated systems.

The distribution of dropped filters for the galaxy and AGN sample are nearly identical.  This indicates that the distribution of the dropped filters is likely a good representation of the distribution of problems in the data rather than being due to systematic problems in the SED models.  In both samples, the most frequently dropped filter is the $Y$-band, which is most likely due to a zeropoint calibration offset, which we have adjusted for on average by examining the SED residuals among optically bright, well-fit sources and applying the average offset between the SED model and the $Y$-band data.  Even with this zeropoint correction, there are still enough objects for which the $Y$-band magnitude is offset from the rest of the SED, and has a small photometric error, that dropping it will cause a significant improvement in the overall SED fit.  The next most commonly dropped filters are the NUV and $U_{S}$ band filters due to the low levels of UV emission from most galaxies.  In the case of AGN, the FUV emission is weak because it is blueward of the Lyman limit for $z\gtrsim0.2$.  The [8.0] and 24\micron\ bands are also frequently dropped for both galaxy and AGN samples.  The distribution of dropped filters for the stellar population is somewhat different from the distribution for galaxies and AGN.  The most commonly dropped filters are the $U_{S}$, \Bw, [3.6], and [4.5] bands.

We limit the total number of dropped filters such that only 1\% of the photometry is excluded from the SED fits.   This leads to 18\% of the sources having a single excluded filter.  For each filter, we only allow up to 1\% of the data for that filter to be dropped.  If more than 1\% of the sources show an improvement greater than $3\sigma$, we drop the 1\% which lead to the greatest \chisq\ improvement.  For a typical source with 15 bands of photometry being fit with the galaxy+AGN models, we require a \deltachisq\ improvement of at least 12.7 due to the exclusion of a single filter.   Different filters can be dropped for the same source depending on whether it is being modelled as a galaxy, AGN, or star/brown dwarf.  However, the total number of bands used in the SED fit must be the same for any given object, regardless of which templates are being used.  The final results are all based on SED fits that exclude this worst 1\% of the photometry.  The choice to exclude 1\% of the photometry is somewhat arbitrary. However, changing the exact choices of these parameters (e.g. increasing the threshold of improvement to $5\sigma$ or excluding a larger or smaller fraction of the data) does not significantly change any of the subsequent results.

We obtain both photometric redshifts and ``bolometric'' luminosities of the individual template components, where ``bolometric'' means the integrated luminosity from 0.03 to 30\micron\ for galaxy templates and 0.1216 to 30\micron\ for the AGN template. The AGN template luminosity is not integrated for wavelengths shorter than Ly$\alpha$ (0.1216 \micron) because the template is not well constrained at these wavelengths due to absorption by the IGM \citep[see][]{Assef2010}.  We note that while the galaxy luminosities are integrated from 0.03 to 30\micron, this is effectively the same as integrating from 0.1216 to 30\micron\ because there is very little far-UV flux in the galaxy templates.

\begin{figure}[!th] 
\centering
\subfigure{\includegraphics[scale=0.5]{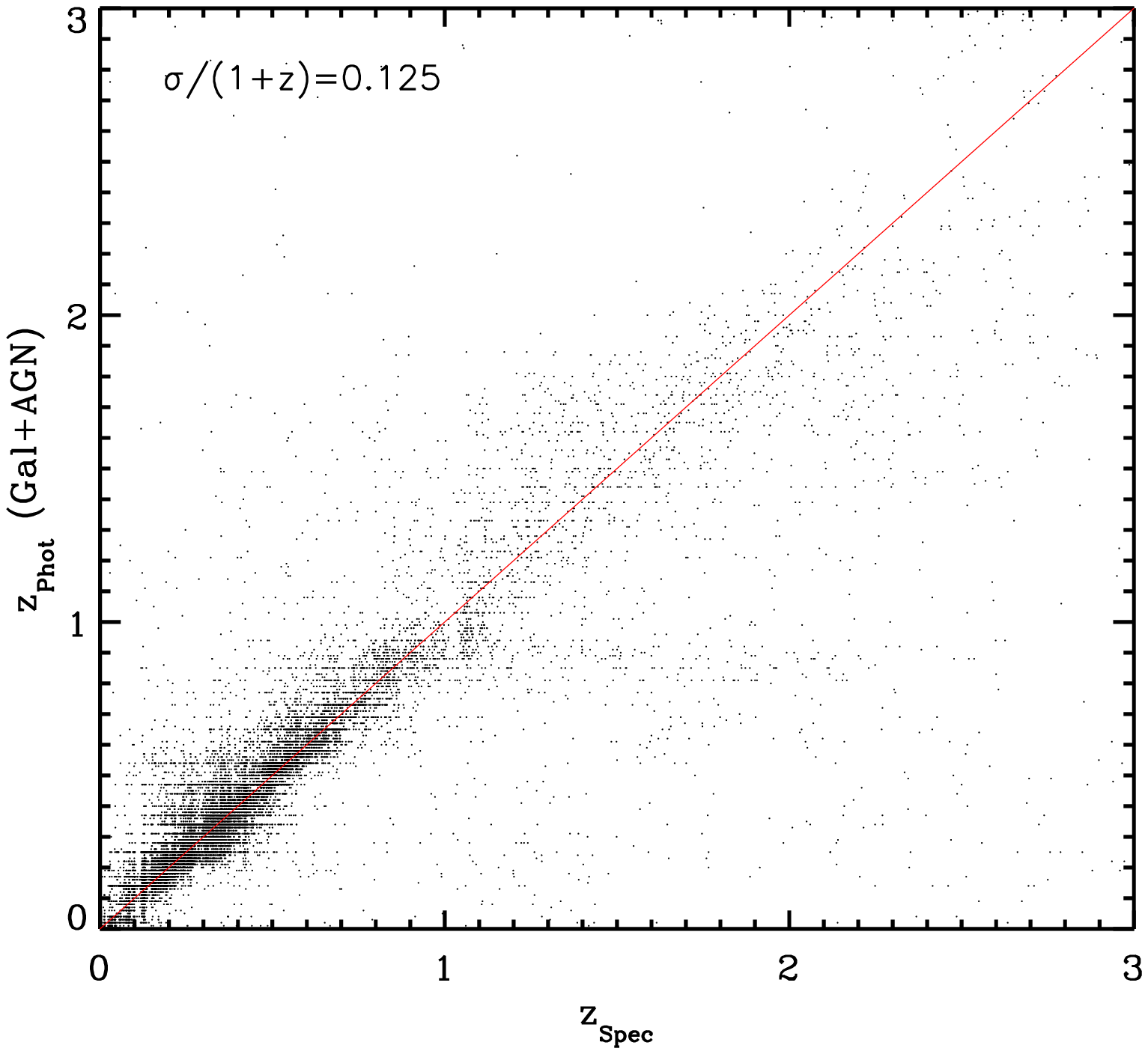}}
\subfigure{\includegraphics[scale=0.5]{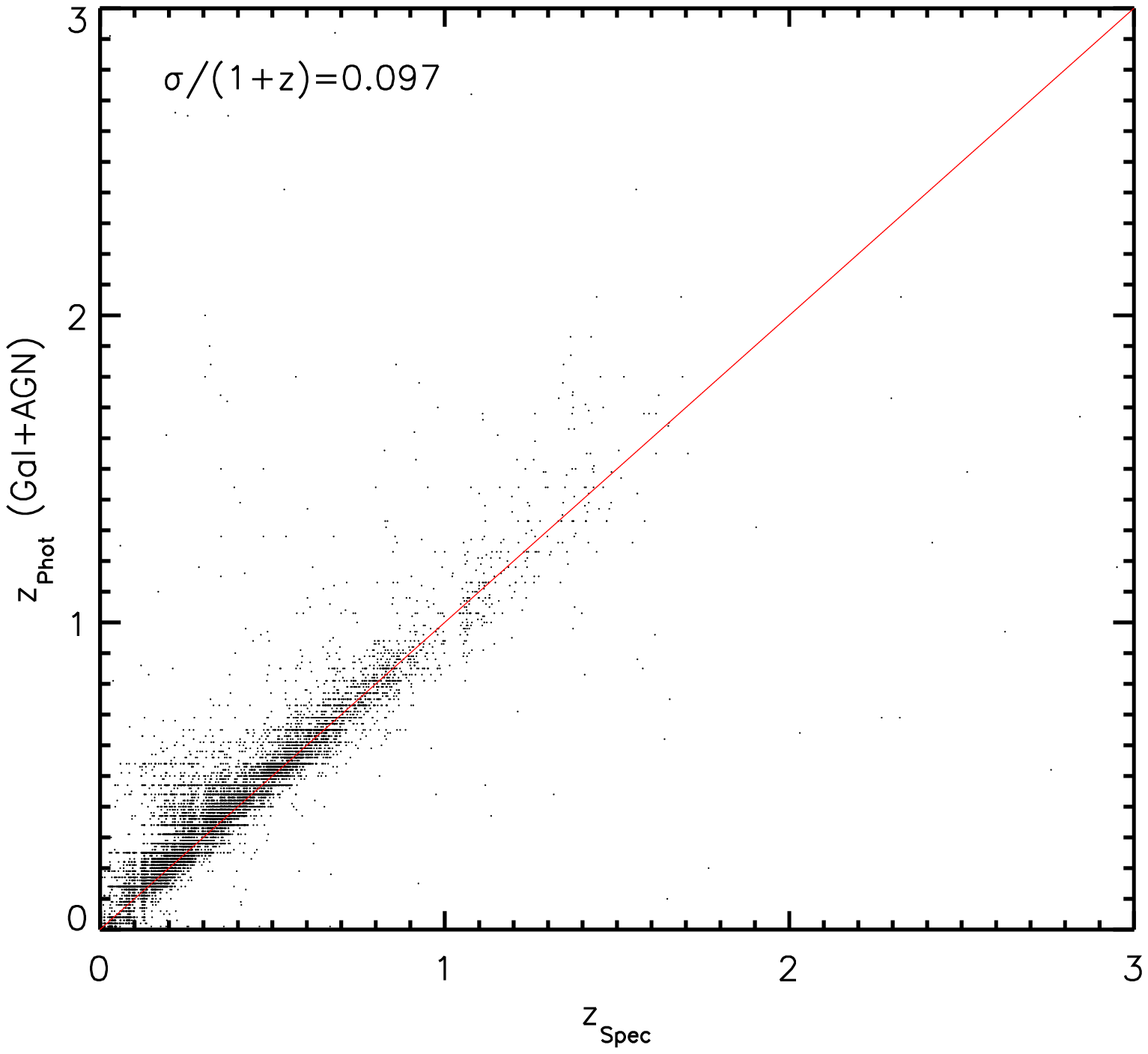}}
\caption{\scriptsize{Left: Photometric versus spectroscopic redshifts for the entire spectroscopic sample in the range $0<z<3$, with the red line denoting a one-to-one relation. The photometric redshift is derived from the galaxy+AGN SED models.  Right: Photometric versus spectroscopic redshifts for sources that are well fit by galaxy-only templates with $F<1$.}} 
\label{fig:photz}
\end{figure}

In Figure~\ref{fig:photz} we compare the photometric and spectroscopic redshifts for a sample of 20,726 galaxies and AGN, primarily from the AGES survey with an additional $\sim$1000 redshifts from Hickox et al. (private communication).   Most of the sources in Figure~\ref{fig:photz} have bolometric luminosities that are dominated by galaxy templates, with 85\% of the galaxies having $<20$\% of an AGN component.  This is consistent with the AGES sample, which is comprised of $\sim$80\% galaxies and $\sim$20\% AGN candidates.  The photometric redshift dispersion is calculated as
\begin{equation}
\frac{\sigma}{(1+z)} =  \sqrt{\frac{1}{N}\sum\limits_{i=1}^N\left(\frac{z_{\rm phot}-z_{\rm spec}}{1+z_{\rm spec}}\right)^{2}}.  
\end{equation}
The photometric redshifts of the entire sample shown in Figure~\ref{fig:photz} have a dispersion of $\sigma/(1+z)=0.125$ and a median offset $<0.001$.  As is typical of photometric redshifts, the dispersion is dominated by the tails of the distribution.  If we drop the worst 5\% of the outliers (largest offsets in $\abs{z_{p}-z_{s}}$), then the dispersion for the remaining 95\% of the sources goes down to $\sigma/(1+z)$=0.061.  Both the dispersion and accuracy of the photometric redshifts are significantly worse for the AGN because AGN-dominated SEDs have few distinct features compared to galaxies \citep{Rowan2008, Salvato2009, Assef2010}.  In many cases, as the AGN component starts to dominate the bolometric luminosity, the SED becomes increasingly flat, leading to a large increase in the photometric redshift uncertainties.

If we separately consider sources that are likely to be AGN-dominated versus those that are galaxy-dominated based on their $F$-ratios, we find that the redshift dispersions are much higher for the AGN-dominated objects, as expected.  We assign the $F<1$ sources to the galaxy sample, and the $F>10$ sources to the AGN sample, thus avoiding the more ambiguous, composite AGN/galaxy sources.  After rejecting the worst 5\% of each sample, the redshift dispersion of the galaxy and AGN samples are  $\sigma/(1+z)=0.040$ and $\sigma/(1+z)=0.169$, respectively.  These dispersions are also consistent with the photometric dispersions of sources targetted as galaxies or AGN candidates in the AGES survey.  The $z_{s}<1$ (primarily galaxies) and $z_{s}>1$ (AGN) have photometric dispersions of $\sigma/(1+z)=0.044$ and 0.204, after rejecting the worst 5\% outliers.  One advantage of our study is that AGES supplies spectroscopic redshifts for most of the luminous AGN where photometric redshifts are most problematic. As we examine AGN that have a stronger host galaxy contribution, the photometric redshifts become increasingly robust.  Our photometric redshift dispersions are comparable to the outlier-excluded dispersions from \citet{Assef2010} for galaxy and point-source AGN of $\sigma/(1+z)=0.041$ and $\sigma/(1+z)=0.184$, respectively.  \citet{Brodwin2006} also calculated photometric redshifts for $\sim$200,000 sources in the \Bootes\ field using a hybrid technique combining a template fitting algorithm with artificial neural nets \citep{Collister2004}, using optical, near-infrared, and IRAC data from NDWFS \citep{Jannuzi1999}, the FLAMEX survey \citep{Elston2006}, and the IRAC shallow survey \citep{Eisenhardt2004} but not the \emph{GALEX}, $U_{S}$, $z$, $Y$, SDWFS, NEWFIRM, or MIPS data included here.  Using the template fitting algorithm, \citet{Brodwin2006} obtained photometric redshift dispersions slightly higher than ours, with $\sigma/(1+z)=0.061$ and 0.341 for the galaxy and AGN samples, respectively.  When \citet{Brodwin2006} use the hybrid technique, their photometric redshift dispersions decrease to  $\sigma/(1+z)=0.047$ and 0.120 for the galaxy and AGN samples, respectively.  

\subsection{Galactic versus Extragalactic Sources}
\label{sec:analysis-galorexgal}

\begin{figure*}[!pht]
\includegraphics[scale=0.85]{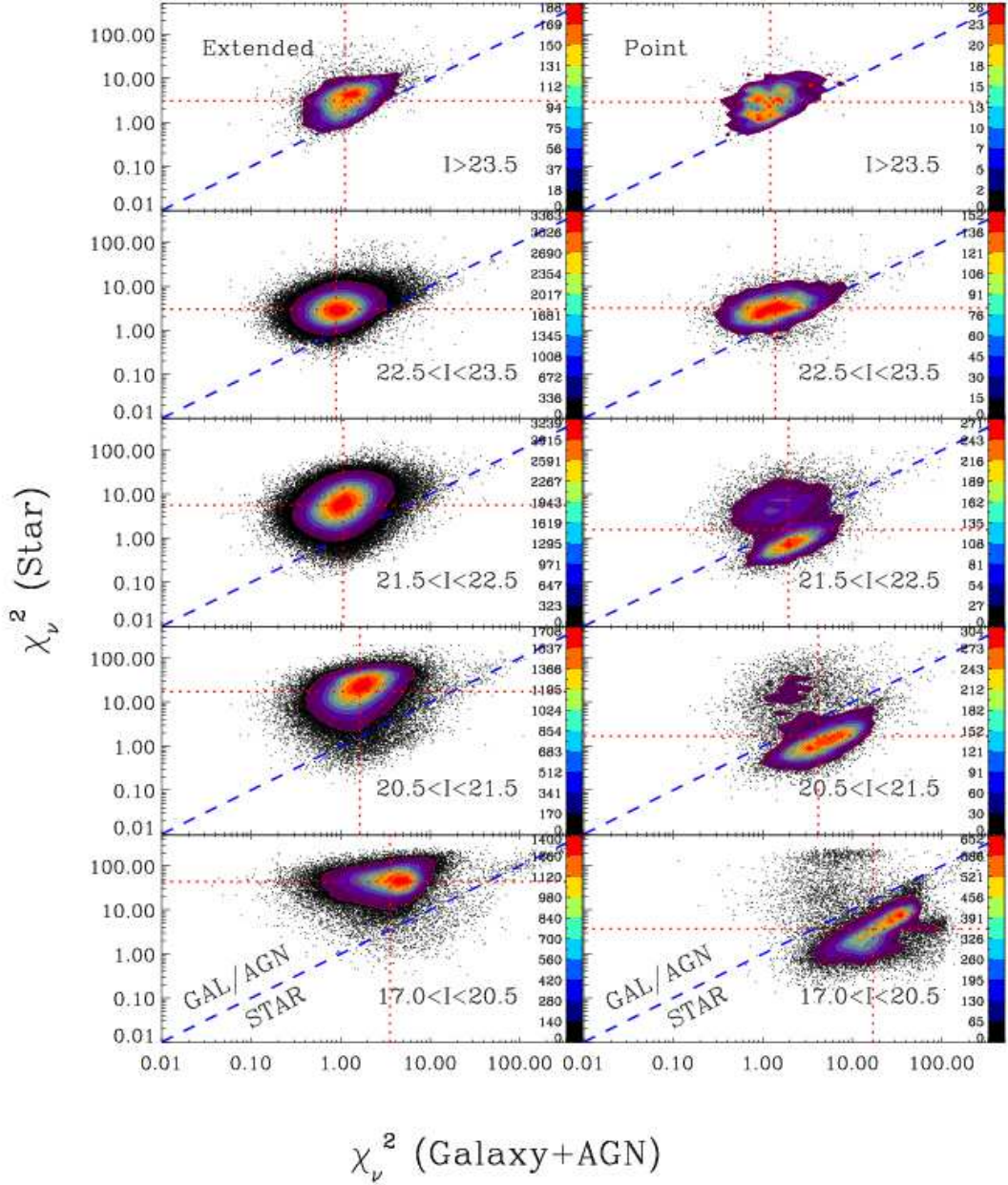}
\caption{\scriptsize{\chisqred\ distributions for the stellar/brown dwarf templates and galaxy+AGN SED models, for extended (left; $S<0.7$) and point-like (right; $S\geq0.7$) sources, in bins of $I$-band magnitude.  Ten density contours are overlaid with the colorscale shown on the right of each panel.  The one-to-one, dashed blue line represents the separation between the stellar/brown dwarf population and the galaxy/AGN population.  The red dotted lines show the median \chisqred\ values for a source with the typical number of degrees of freedom.}}
\label{fig:chi2}
\end{figure*}

An important step in sorting through the different types of sources in the \Bootes\ field is to separate the Galactic from the extragalactic sources.  To do this, we examine the \chisqred\ of the stellar templates as compared to AGN+galaxy templates, as shown in Figure~\ref{fig:chi2}. The sample is split into extended and point-like sources, and further separated into bins of $I$-band magnitude.   Extended and point-like source definitions are determined from the SExtractor stellarity index $S$, which has values from 1 (point source) to 0, measured on the NDWFS $I$-band images.  We consider sources with $S<0.7$ as extended and $S\geq0.7$ as point-like because there is a minimum in the distribution of stellarity indices at $S\sim0.7$.   Based purely on \chisqred, sources above the blue diagonal line have SEDs which are better fit by the galaxy+AGN templates, while sources below the line are better fit as stars or brown dwarfs.  We will refer to  those  regions in \chisqred\ space as the ``extragalactic'' and the ``stellar'' zones, respectively.  Because the distinction between stars and extragalactic sources is not a continuum, we do not use $F$ statistics in this separation.  

As expected, most of the extended sources shown in Figure~\ref{fig:chi2} (left) lie above the blue line, indicating that they are better fit by the galaxy and AGN templates than the stellar or brown dwarf templates.  In the optically brightest bin ($17<I<20.5$), nearly all the sources lie above the blue line and are better fit as extragalactic sources. As the sources become optically fainter, the cloud moves to smaller values of  both \chisqred(Gal+AGN) and \chisqred(Star) as the photometric uncertainties increase, hitting  minimum \chisqred\ values in the $22.5<I<23.5$ bin.  In the optically faintest $I>23.5$ bin, \chisqred\ increases slightly, reflecting the increased uncertainty in the SED fits due to fewer available high S/N data points.   In the brightest bins, systematic errors  (e.g., offsets between filter zeropoints, photometric anomalies, problems in the templates) dominate the true total uncertainty in the SED fits rather than the photometric errors, leading to the high values of \chisqred.  In the $17.0<I<20.5$ bin, the extended and point-like sources have median \chisqred\ values of $\langle$\chisqred(Gal+AGN)$\rangle\simeq3.5$ and $\langle$\chisqred(Star)$\rangle\simeq3.5$.  In comparison, the \chisqred\ of extended and point-like sources in the fainter $21.5<I<22.5$ bin is lower, with $\langle$\chisqred(Gal+AGN)$\rangle\simeq1.0$ and \chisqred(Star) $\simeq$ 1.8.   Note that the $\langle$\chisqred(Star)$\rangle\simeq1.8$ of $21.5<I<22.5$ point-like sources is misleadingly high due to the presence of contaminating galaxies and AGN.  If we consider only the point-like sources that are best fit by stellar templates in the $21.5<I<22.5$ bin, \chisqred(Star) decreases to $\simeq$ 0.8.    The overall shifts of \chisqred\ suggest that the photometric error bars at the brightest magnitudes should be broadened by closer to 0.10 mag rather than 0.05 mag in order to encompass systematic uncertainties.

The right side of Figure~\ref{fig:chi2} shows the same sequences for point-like sources.  There are three classes of point-like sources in Figure~\ref{fig:chi2}: Galactic stars and brown dwarfs, unresolved high-luminosity AGN, and galaxies misclassified as point-like sources due to their apparent compactness.  Among the optically bright, point-like sources, there is an elongated horizontal cloud of sources at \chisqred(Star)$\sim100$, which is mainly populated by bright AGN.  The remainder of the sources in the Galaxy/AGN region are largely comprised of lower luminosity AGN with a more significant host component.  Hosts generally have more ``stellar'' SEDs leading to lower values of \chisqred(Star) than for the high-luminosity AGN.    However, the dominant population of optically bright point-like sources are clearly best fit by stellar SEDs.  While the \chisqred(Star) are sometimes high for these sources, it is not driven by saturation in the $17.0<I<20.5$ magnitude range, but rather due to the simple suite of stellar template models (e.g. only solar metallicity) and to the sometimes overly small photometric uncertainties.

\begin{figure*}[!t]
\centering
\subfigure{\includegraphics[width=3in]{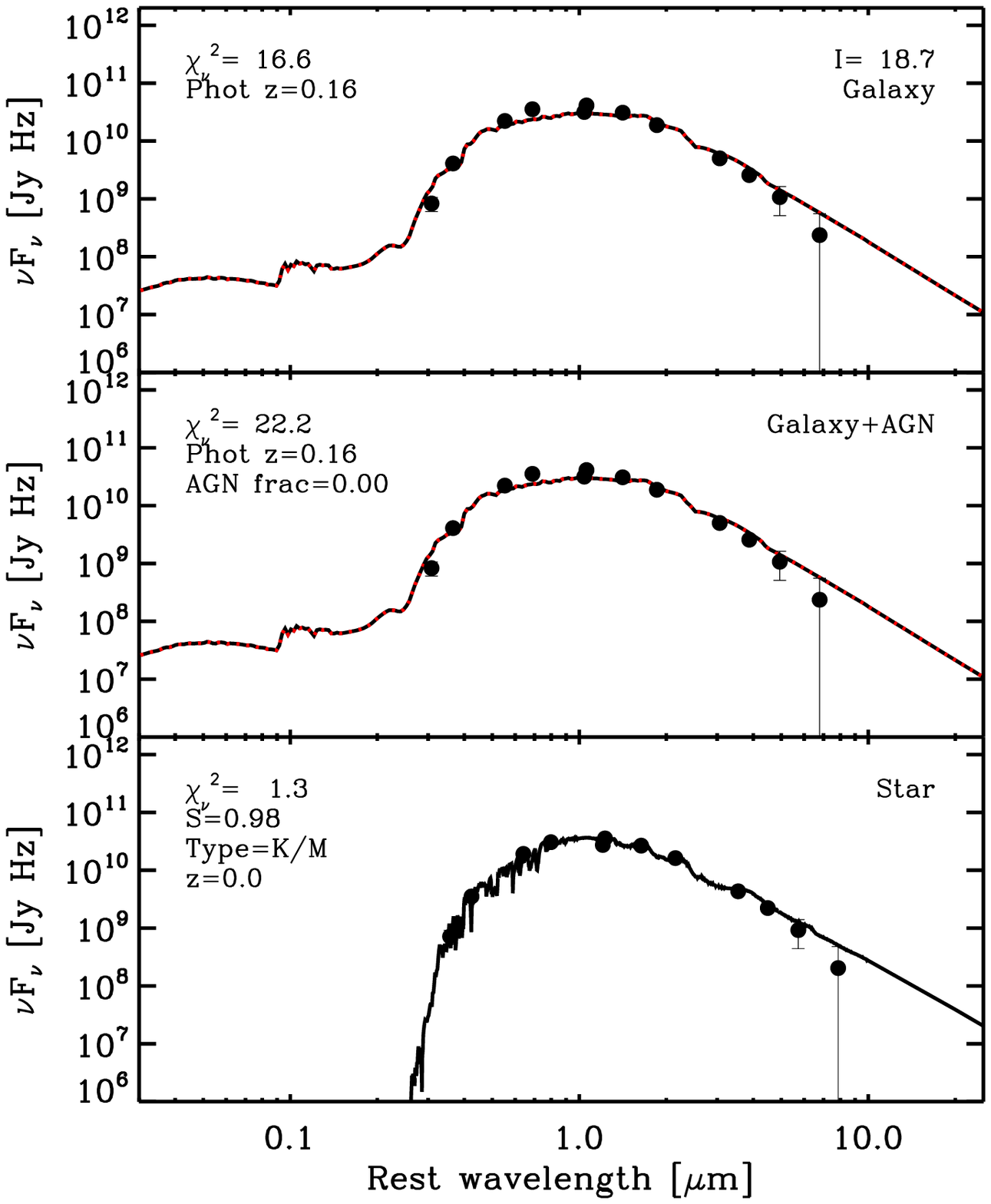}}
\subfigure{\includegraphics[width=3in]{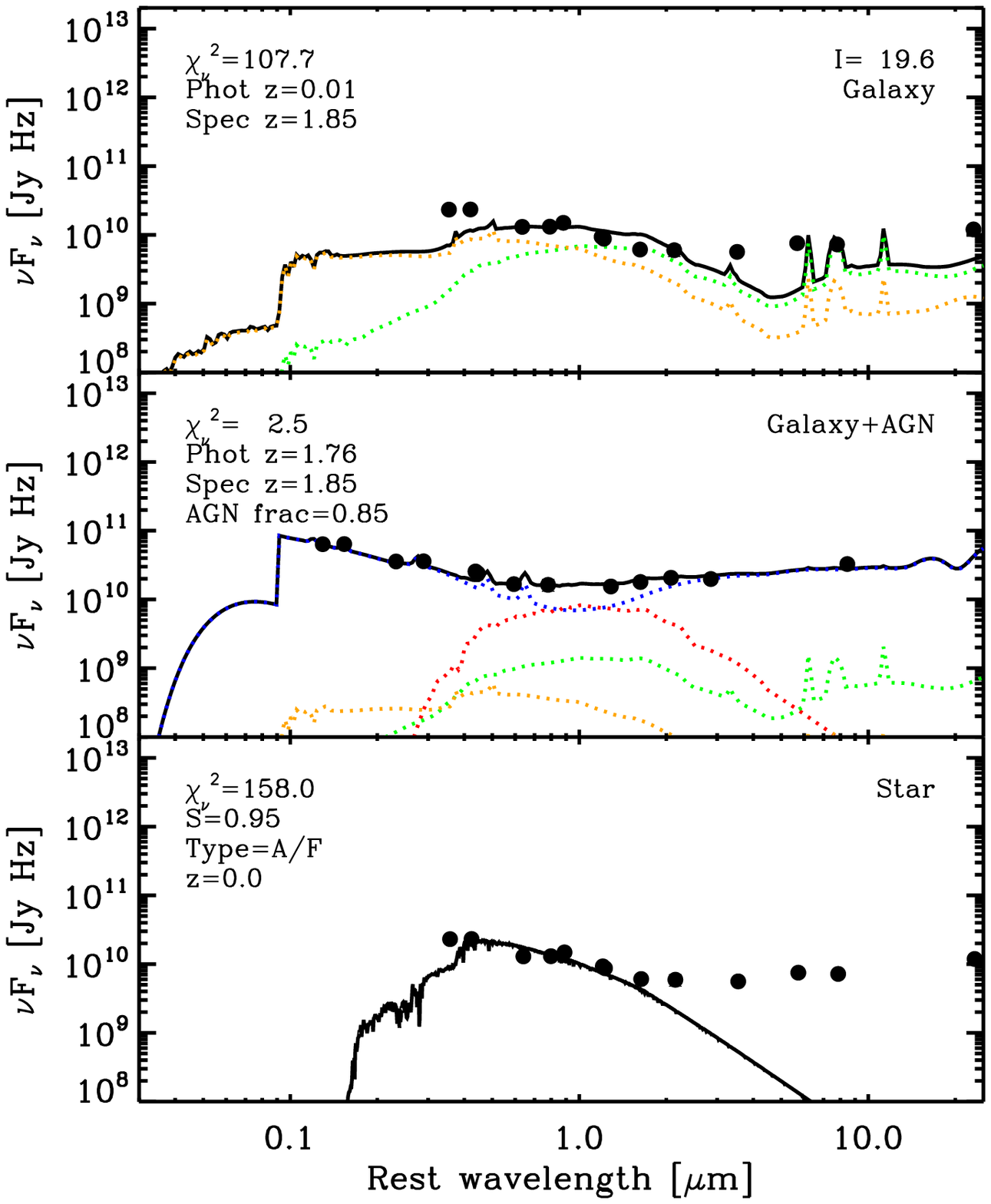}}
\caption{\scriptsize{Left: SED of a K/M star fit with galaxy (top), galaxy+AGN (middle) and stellar (bottom) templates.  Note that while the \chisqred\ has increased for the galaxy+AGN model, this is only due to the change in $\nu$.  Right: SED of a known optically bright, point-like, X-ray AGN. }}
\label{fig:SED_optbright}           
\end{figure*}

Figure~\ref{fig:SED_optbright} shows the SEDs of two optically bright point-like sources.  The source on the left is clearly best fit by a stellar template.  On the right is a point-like, luminous X-ray source, best fit by an AGN.   The SED of a cool star has a shape that is somewhat similar to the SED shape of an elliptical galaxy longward of the UV wavelengths.  Nonetheless, in most cases, the fits are able to easily distinguish between stars and elliptical galaxies.  The \chisqred\ of the star shown in Figure~\ref{fig:SED_optbright} (left) is significantly better fit with a stellar model (\chisqred$\sim1$) than a galaxy or galaxy+AGN model (\chisqred$\sim20$).   The SEDs of luminous AGN are very non-stellar (more so than the SEDs of non-AGN galaxies), which means they are  well separated from the stellar regions of Figure~\ref{fig:chi2}.  The  SED shown in Figure~\ref{fig:SED_optbright} (right) is an example from the elongated horizontal cloud of sources  at \chisqred(Star)$\sim100$.   The photometric redshift of this AGN is $z_{p}=1.76$ and the measured spectroscopic redshift is $z_{s}=1.85$, confirming that the SED fit has correctly fit the source as AGN.  As we consider sources with less luminous AGN and a stronger host galaxy component, the composite SED becomes more ``star-like'' in structure and therefore less separated in \chisqred\ space from the stellar regions.

As the point-like sources become optically fainter, the \chisqred\ distribution becomes bi-modal between the stellar and the extragalactic regions of the figure, as seen in the $20.5<I<21.5$, $21.5<I<22.5$, and $22.5<I<23.5$ panels.  The point-like sources that fall into the galaxy/AGN region of \chisqred\ space are a mix of point-like AGN and compact galaxies with a low stellarity index.  In these three magnitude bins, $\sim$13\%, 17\%, and 24\% of the extragalactic SEDs have $F>3$ for the galaxy+AGN fits over the galaxy-only fits.  This nominally corresponds to a $\sim$90\% probability that an AGN component is required modulo the caveats discussed in \S\ref{sec:analysis-SEDfitting}.  As we examine fainter sources, the morphological star-galaxy separation fails more frequently and as a result, the distributions of \chisqred\ for extended and point-like sources look increasingly similar in the $22.5<I<23.5$ and $I>23.5$ panels.  Finally, by the time we reach the optically faintest bin, the distribution of \chisqred\ for point-like sources looks nearly identical to that of extended sources.  The vast majority of these are extragalactic, and the stellar population is naturally excluded in this magnitude bin due to the criteria that optically faint $I>23.5$ sources are required to have mid-IR detections with $[4.5]<18$.

\begin{figure}[!t]
\centering
\includegraphics[width=3in]{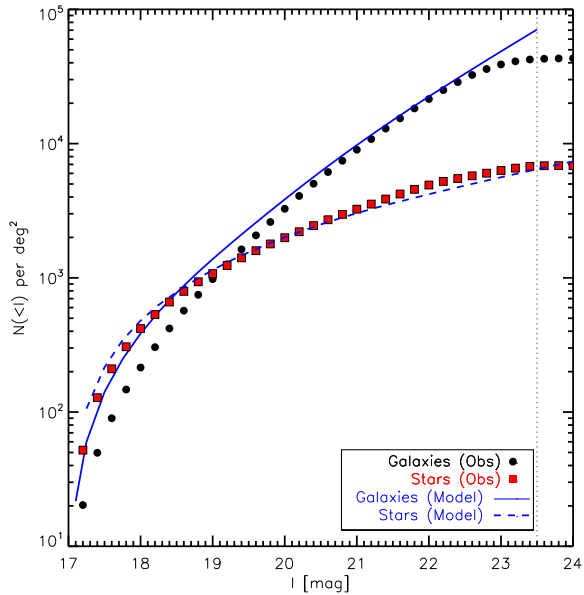}
\caption{\scriptsize{Integrated number of expected and observed ($I>17$) stars and galaxies as a function of $I$ magnitude. The dotted vertical line indicates the faint optical limit of our sample.  Since we are using aperture magnitudes, the observed galaxy counts fall below the model at bright magnitudes.}}
\label{fig:starcounts}           
\end{figure}

Figure~\ref{fig:starcounts} shows the expected and observed integrated number counts of the stellar and extragalactic populations.  The expected number counts of galaxies are calculated using \citet{Ellis2007} and the stellar number counts are from the  Besan\c{c}on stellar population synthesis model \citep{Robin2003}.  The observed counts simply use the \chisqred\ separation illustrated in Figure~\ref{fig:chi2}.  The resulting number counts for both the galaxy and stellar populations match the models quite well, although the completeness of the galaxy sample begins to drop rapidly for $I\gtrsim23$.  The mismatch for bright galaxies arises from using fixed 6\farcs0 diameter aperture magnitudes rather than integrated Kron magnitudes.  This underestimates the luminosities of the brighter galaxies and shifts the observed number counts to lie to the left of the models.

\subsubsection{Outliers and Degeneracies}

As discussed above and shown in Figure~\ref{fig:chi2}, the \chisqred\ separation of extragalactic and Galactic sources works well for the vast majority of our sample.  However, as with any large sample based on automatic fits to data, there are bound to be some failures.  In some cases this is due to a combination of inadequate data, and/or degeneracies between the galaxy and stellar templates, while in other cases it is due to one or more deviant data points that skew the fit.

\begin{figure*}[!pht] 
\centering
\includegraphics[scale=0.85]{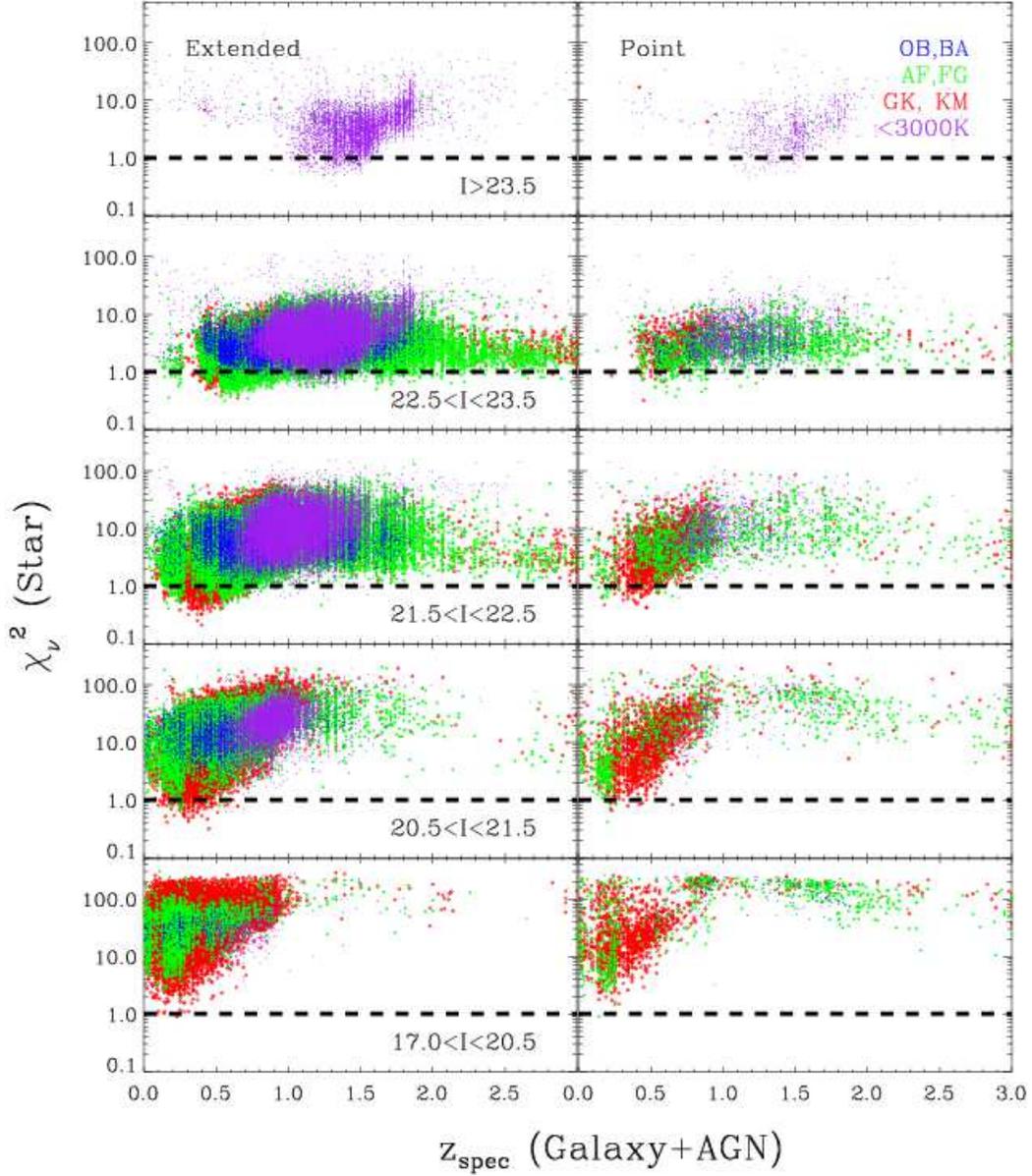}
\caption{\scriptsize{Distribution of non-stellar sources (as defined by the \chisqred\ separation in Figure~\ref{fig:chi2}) for the best fit stellar/brown dwarf template as a function of photometric redshift from the galaxy+AGN SED models, separated into extended (left) and point-like (right) sources, in bins of $I$-band magnitude.   The color-coding indicates which of the stellar/brown dwarf templates produced the lowest \chisqred\, roughly split into ``hot'' (O/B and B/A), ``intermediate'' (A/F, F/G), ``cool'' (G/K, K/M), and ``cold'' ($<3000$K) ``stars'', in blue, green, red, and purple, respectively.  } }
\label{fig:photz_stellartype}
\end{figure*}

Figure~\ref{fig:photz_stellartype} explores possible degeneracies between the stellar templates and galaxies by showing the distribution of \chisqred(Star) as a function of the photometric redshift for the galaxy+AGN SED models as applied to the non-stellar sources (anything classified as a galaxy or AGN based on the \chisqred\ of their SED fits; see Figure~\ref{fig:chi2}).  We do not include objects better fit as stars or brown dwarfs in Figure~\ref{fig:photz_stellartype}.  The data are again split into extended (left) and point-like (right) sources and then further into $I$-band magnitude bins.  The different colors represent best fit stellar/brown dwarf templates split into groups by temperature.  Sources below the horizonal line at \chisqred(Star)=1.0 are galaxies or AGN that are nominally well-fit by a stellar template.

Among the optically bright galaxies and AGN, the best fitting stellar templates are overwhelmingly the cool stars, with some contribution from intermediate temperature stars, essentially because the SEDs of hotter stars peak at too short a wavelength.  There is also a well-defined lower envelope to the \chisqred(Star) and $z_{p}$(Gal+AGN) distribution for the optically brighter galaxies and AGN.  As the galaxy/AGN redshift increases, the \chisqred(Star) values rise sharply -- optically bright $z\gtrsim0$ galaxies and AGN are poorly fit by all stellar/brown dwarf templates.

As sources become fainter than $I\sim21.5$ (top three panels), there tends to be more of a degeneracy between galaxy and stellar templates.  This is reflected by the increasing number of sources for which stellar templates can produce reasonable values of \chisqred$\sim1$.  The range of galaxy redshifts that are compatible with stellar/brown dwarf templates generally increases with decreasing optical magnitude (except in the case of $I>23.5$, due to the [4.5] constraint).  Also, as the galaxies become optically fainter, the cool star/brown dwarf templates (purple) become more prevalent, and overlap with higher redshift galaxies as well as for a wider range of galaxy redshifts.  The intermediate and cool star templates (green and red) can fit some galaxies  with a range of redshifts from $0.1\lesssim z_{p}\lesssim2.0$ with reasonable \chisqred\ values in the $21.5<I<22.5$ magnitude bin.  In the $22.5<I<23.5$ bin, the intermediate and cool stellar templates that are fixed at $z=0$, can fit galaxies in almost the entire range of photometric redshifts and still yield \chisqred\ values close to 1.  In the $I>23.5$, most sources are best fit with the \Teff$<3000$K templates.  We did attempt to identify brown dwarfs using these fits but it works poorly because they are usually detected in only a few bands.

\section{The Search for AGN}
\label{sec:AGNcandidates}

While AGN emit energy in a broad range of wavelengths, most existing AGN selection techniques utilize only a narrow slice of the AGN spectrum such as the sources' optical or mid-IR color.  Each of these AGN diagnostics is sensitive to a particular type or ``viewing angle'' in the unified picture of AGN and the balance between AGN and host emission \citep[e.g.][]{Hickox2009, Mendez2013}.  However, we have the advantage of having up to 17 bands that span the wavelength range from the near-UV to the mid-IR, so fitting galaxy and AGN templates to these data should allow for  more complete AGN selection than using only a few colors within a limited wavelength window.

\begin{figure*}[!tph] 
\centering
\includegraphics[scale=0.8]{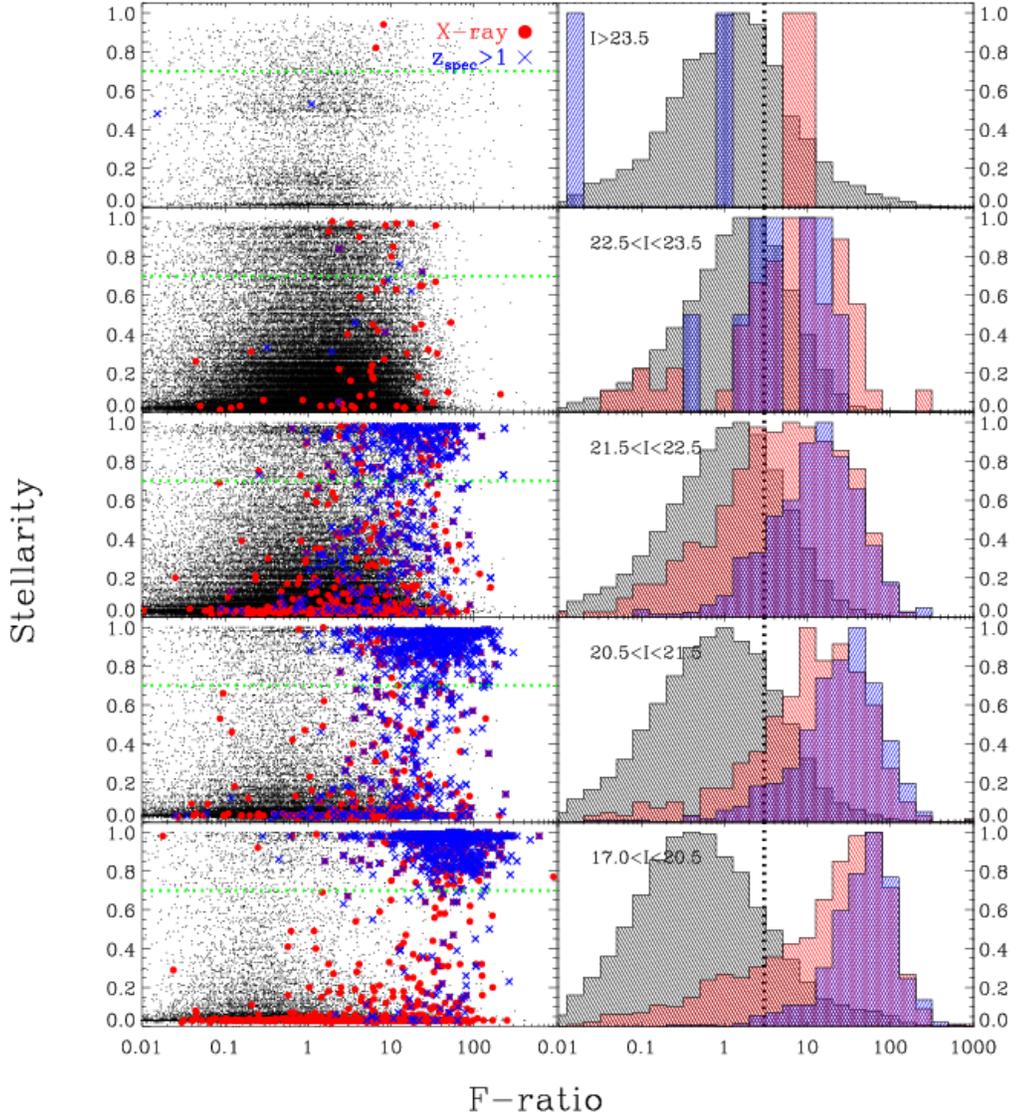}
\caption{\scriptsize{The left panels show stellarity ($S$) as a function of the $F$-ratio between the galaxy and galaxy+AGN models, in bins of $I$-band magnitude, excluding sources that are best fit by the stellar templates.  The dotted horizontal line (left) at $S=0.7$ shows our adopted separation between extended ($S<0.7$) and point-like ($S\geq0.7$) sources.  The X-ray and $z_{s}>1$ samples are shown using red circles and blue crosses, respectively.  The right panels show the normalized distribution of $F$-ratios for the full, X-ray, and $z_{s}>1$ samples in black, red, and blue histograms, respectively.  The dotted vertical line at $F=3$ is meant to provide only a visual aid, and corresponds to $\sim$90\% AGN probability based purely on the $F$-ratio distribution (see caveats in \S\ref{sec:analysis-SEDfitting}.  Note that the $F$-ratio distributions for the X-ray and $z_{s}>1$ samples peak at larger $F$-ratios in all magnitude bins with the exception of the faintest bin, where there are only 2 X-ray and 2 $z_{s}>1$ sources.}} 
\label{fig:fratio_stellarity}
\end{figure*}

Figure~\ref{fig:fratio_stellarity} shows the distribution of sources in stellarity as a function of the $F$-ratio between the galaxy and galaxy+AGN templates, in bins of $I$-band magnitude for all sources except those classified as stars based on the \chisqred\ separation shown in Figure~\ref{fig:chi2}.  The horizontal line at $S=0.7$ is the division we use between extended ($S<0.7$) and point-like ($S\geq0.7$) sources.   The vertical line at $F=3$ corresponds to a $\sim$90\% probability for the existence of an AGN component in the SED for the typical number of degrees of freedom (see \S\ref{sec:analysis-SEDfitting}).  As discussed earlier,  $z_{s}>1$  sources are all AGN, so we use this as one comparison sample.  As a second comparison sample we use  X\Bootes\ sources with at least 4 X-ray counts.
 The right panels of Figure~\ref{fig:fratio_stellarity} show the distribution of $F$-ratios for the entire non-stellar sample (galaxies and AGN) in black, and the X-ray and $z_{s}>1$ AGN non-stellar samples in red and blue, respectively, where ``non-stellar'' is purely defined by having \chisqred(Galaxy+AGN)$<$\chisqred(Star).  The panels are again separated into $I$-band magnitude bins, and the histograms are normalized to a peak of unity.  The two AGN samples clearly tend to reside at higher $F$-ratios relative to the entire galaxy/AGN population, particularly for the brightest optical magnitudes.

Among the brightest sources, the $z_{s}>1$ $F$-ratio distribution is roughly log normal and centered at $F\sim50$, with most of the contribution coming from point-like sources.  The X-ray sample similarly peaks at F$\sim$40, but with a much broader, assymetric distribution that extends towards low $F$-ratios, presumably because X-ray selection is more sensitive to host-dominated, spatially extended AGN than the selection methods that dominate the $z_{s}>1$ AGN sample \citep{Hickox2009,Mendez2013}.  However, for the $I\gtrsim20$ AGN candidates in AGES, there are complex sampling biases.  For example, point-like sources were selected more broadly than extended sources (e.g., all 24\micron\ quasars were required to be optically point-like -- see \citealt{Kochanek2012}).  These sampling bias effects will be minimized for the $I<20.5$ bin, and even here the X-ray sample shows a broader $F$-ratio distribution.  It is unlikely to be an effect of obscuration since mid-IR SEDs are fairly immune to moderate levels of dust ($\mathrm{A_{v}}\lesssim$10, \citealt[see][]{Assef2011}).

The only non-AGN that should appear in the X-ray samples are very low redshift galaxies, where X-ray emission by the integrated binary populations can dominate, and X-ray active stars.  Out of the 136 X-ray sources with $F<1$ in the optically bright bin of Figure~\ref{fig:fratio_stellarity}, 113 have spectroscopic redshifts.  Among these, nine sources have $z_{s}<0.1$, four of which are stars ($z=0$) and five which have SEDs that are consistent with star-forming galaxies or host-galaxy-dominated AGN.   We also checked for a correlation between the off-axis angle of the source in the X\Bootes\ observations and the $F$-ratios of the optically bright and extended sample.  Due to the increasing size of the \emph{Chandra}/ACIS point spread function (PSF) with distance from the center of the image, X-ray detections made at larger off-axis angles have an increased likelihood of being matched to the wrong optical counterpart.  However, we find no relation between the low $F$-ratio X-ray sources and their X-ray off-axis angles, indicating that there are likely to be few false positives, and the tail of low $F$-ratio X-ray sources in Figure~\ref{fig:fratio_stellarity} is primarily from extended host-galaxy-dominated AGN.  The spatially extended X-ray AGN tend to have higher $F$-ratios than the overall galaxy sample in all magnitude bins.

The left panels of  Figure~\ref{fig:fratio_stellarity} also show that as the sources become optically faint, the median $F$-ratios of the point-like sources and the extended sources start to converge.  In the optically brightest bin, the point-like sources are clustered at high $F$-ratios, while the extended sources cluster at significantly lower $F$-ratios. Yet in the faintest bin, the distinction between the extended and point-like sources in terms of their $F$-ratio values is less clear, presumably due to contamination by compact and/or high-redshift galaxies.  In the fainter bins, we also have fewer X-ray and $z_{s}>1$ sources due to the X-ray and redshift survey flux limits.

\begin{figure*}[!tph] 
\centering
\includegraphics[scale=0.8]{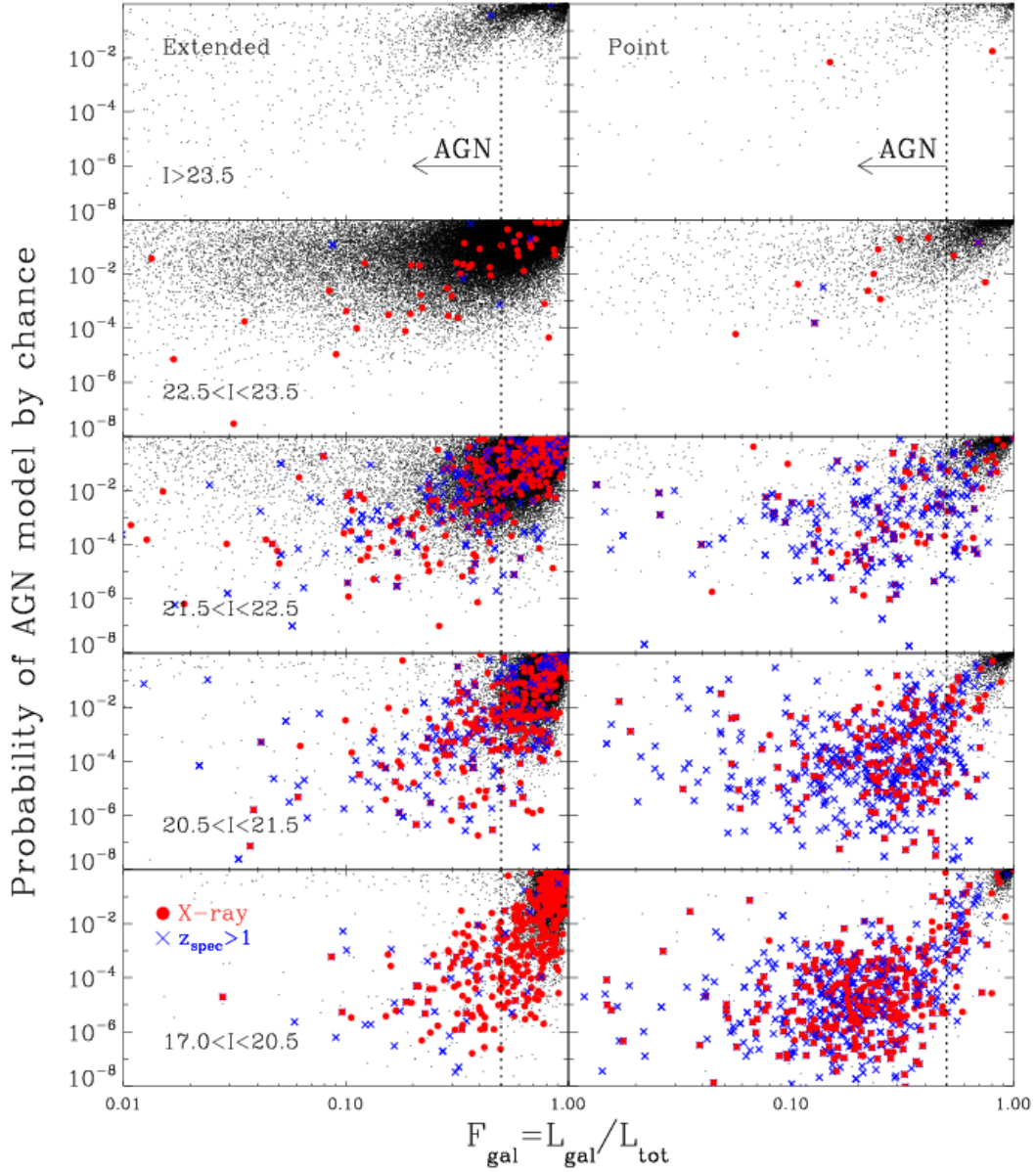}
\caption{\scriptsize{$F$-test probability for the significance of an AGN component as a function of host galaxy luminosity fraction ($L_{\rm gal}/L_{\rm total}$) for extended ($S<0.7$, left) and point-like ($S\geq0.7$, right) extragalactic sources in bins of $I$-band magnitude.  Small probabilities mean that the improvement in the fit from adding the AGN component is unlikely to occur by chance.  The vertical dotted line at  $L_{\rm gal}/L_{\rm tot}=0.5$ marks where the AGN and host contribution are equal, with increasing AGN fraction in the direction of the arrow.  The X-ray and $z_{s}>1$ samples are shown by the red circles and blue crosses, respectively.}}
\label{fig:Pvalue_Lgal}
\end{figure*}

Figure~\ref{fig:Pvalue_Lgal} shows the $F$-test probability for the galaxy and galaxy+AGN models as a function of the host galaxy luminosity fraction $F_{\rm gal}=L_{\rm gal}$/$L_{\rm total}$ for extended (left) and point-like (right) sources in bins of $I$-band magnitude.  Note that \citet{Assef2010} found that $F_{\rm gal}$ was relatively accurate even for sources with poor photometric redshift estimates. As explained in \S\ref{sec:analysis-SEDfitting}, low probabilities imply that the AGN component is unlikely to be improving the fits by chance.    The dotted vertical line at $F_{\rm gal}$=0.5 indicates where the luminosity contribution from the AGN is equal to the contribution from the host galaxy, with the arrow pointing in the direction of increasing AGN fraction.  The X-ray and $z_{s}>1$ samples are overplotted in large red and blue filled circles, respectively.  We will refer to AGN candidates with $F_{\rm gal}=L_{\rm gal}$/$L_{\rm total}>0.5$  as ``host-dominated''.

As expected, we see that the extended-source AGN tend to be more host-dominated than their point-source counterparts.  This is reassuring since the SED fits have no knowledge of the morphological information.  Among the X-ray AGN sample, 74\% of extended AGN are host-dominated, while only 23\% of point-source AGN are host-dominated.  For the $z_{s}>1$ AGN sample, 33\% and 10\% of  extended and point-source AGN are host-dominated, respectively.  The difference between extended and point source AGN is less dramatic for the $z_{s}>1$ sample compared to the X-ray sample, but this again cannot be trivially interpreted because morphology played a role in the selection process. Not surprisingly, there is a strong correlation between the $F$-test probability for the significance of an AGN component and the AGN luminosity fraction ($F_{\rm AGN}=1-F_{\rm gal}$) at bright magnitudes.  In this regime ($I<21.5$), it should be straightforward to select AGN, even when the total luminosity is dominated by the host galaxy.  This continues to be true for point-like sources until $21.5<I<22.5$, but in the fainter bins it is unclear how well such a selection method would work given the available photometry.

\subsection{Comparison to mid-IR and optical color selection} 

\label{sec:compare}

In the previous sections, we have shown that there is significant overlap between known samples of X-ray and $z_{s}>1$ AGN to sources selected as AGN candidates based on the SED fitting and $F$-ratios.  In this section, we will examine how the AGN sample selected from SED fitting compares to mid-IR color and optically selected AGN samples.

\subsubsection{Mid-IR}
                         
As discussed in \S\ref{sec:Intro}, the mid-IR SEDs of AGN are distinct from the SEDs of normal galaxies due to emission from the accretion disk (higher redshifts) or dust (lower redshift).  If the luminosity of the AGN is at least comparable to that of the host galaxy, the composite SED in the infrared will resemble a power-law.  Thus, mid-IR colors can be used to identify AGN candidates depending on the depth of the IR data and the relative strength of the AGN and the host component.  In this section we investigate how the $F$-ratio values compare to the ``standard'' mid-IR color selections of \citet{Lacy2004}, \citet{Stern2005}, and \citet{Donley2012}.

\begin{figure*}[!t] 
\centering
\subfigure{\includegraphics[width=3in]{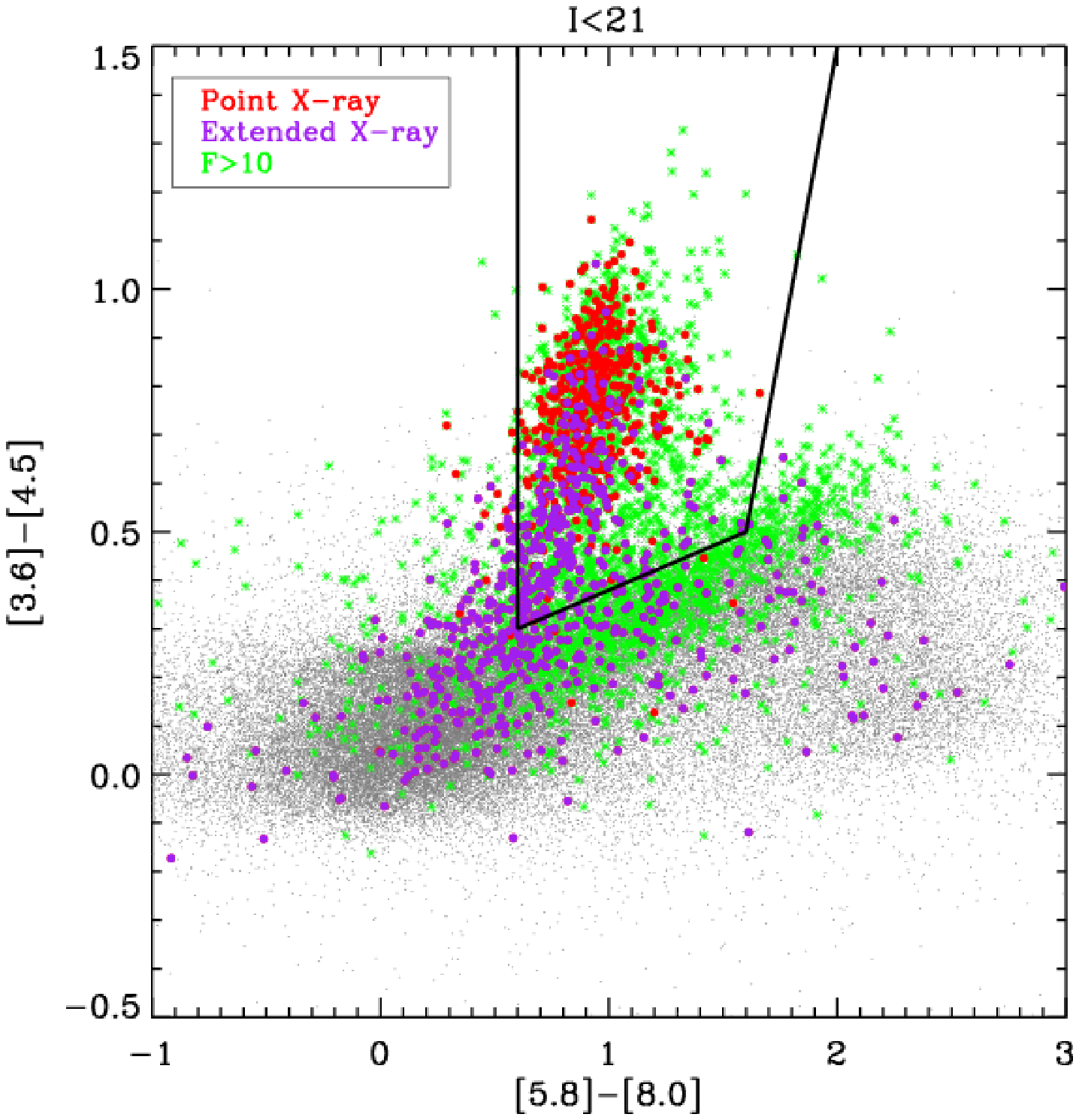}}
\subfigure{\includegraphics[width=3in]{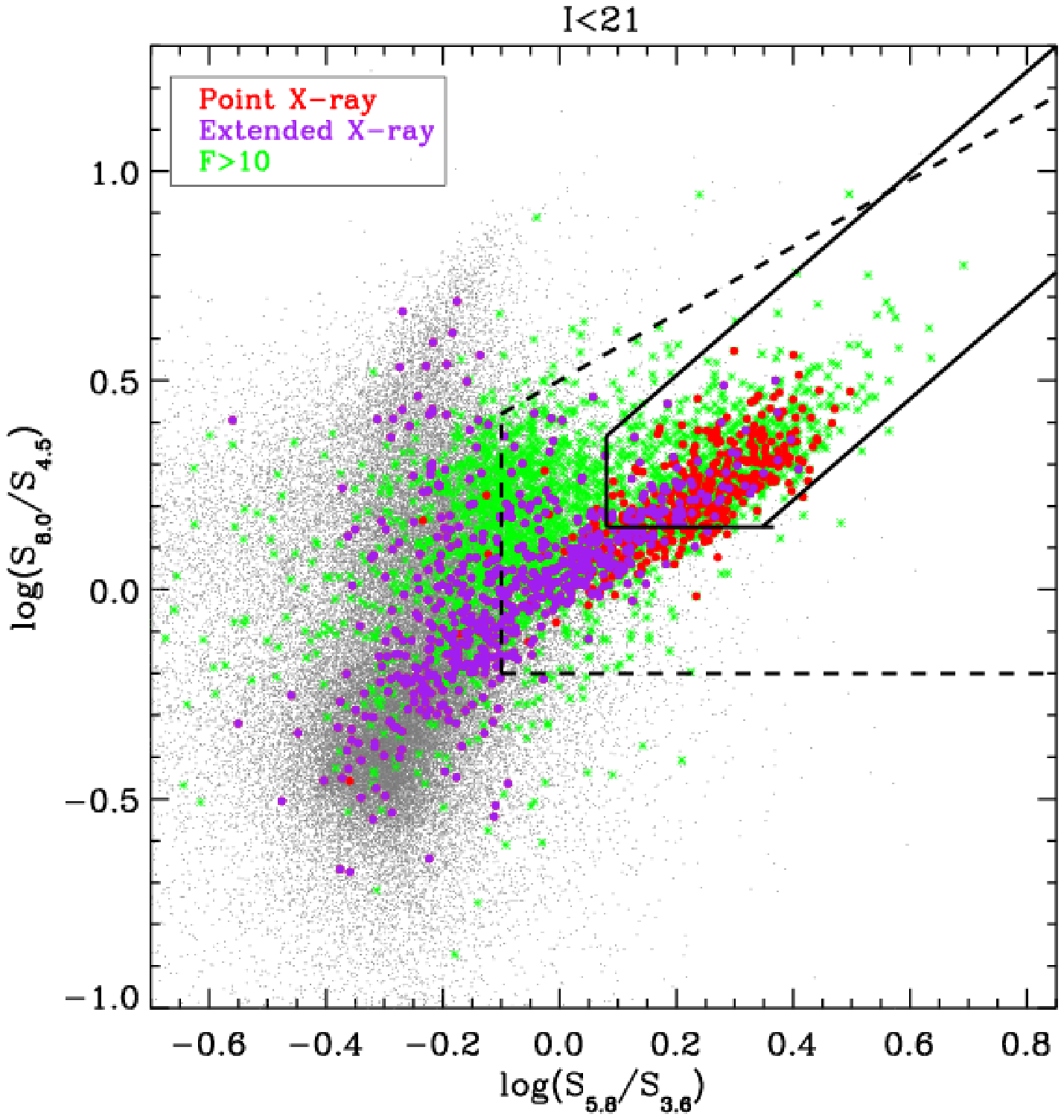}}
\caption{\scriptsize{IRAC color-color diagram with X-ray and $F>$10 sources overplotted in red and green symbols, respectively.  Left: IRAC color criteria from \citet{Stern2005} shown in solid black.  Right: IRAC color criteria from \citet{Lacy2004} and \citet{Donley2012} shown in dashed and solid black lines, respectively.  The samples are limited to sources with $I<21$. }}
\label{fig:stern_wedge}
\end{figure*}

\begin{figure*}[!t] 
\centering
\subfigure{\includegraphics[width=3in]{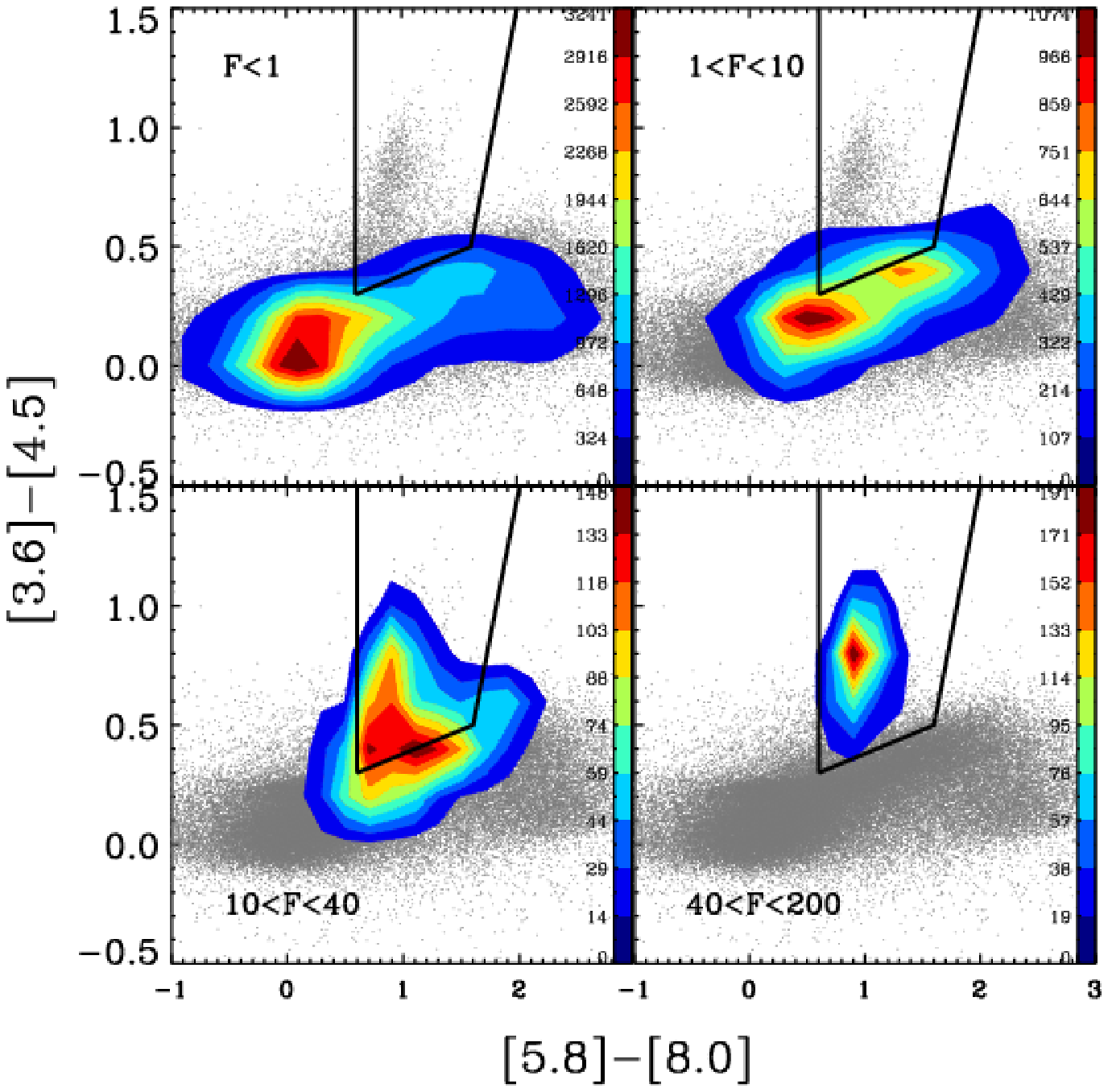}}
\subfigure{\includegraphics[width=3in]{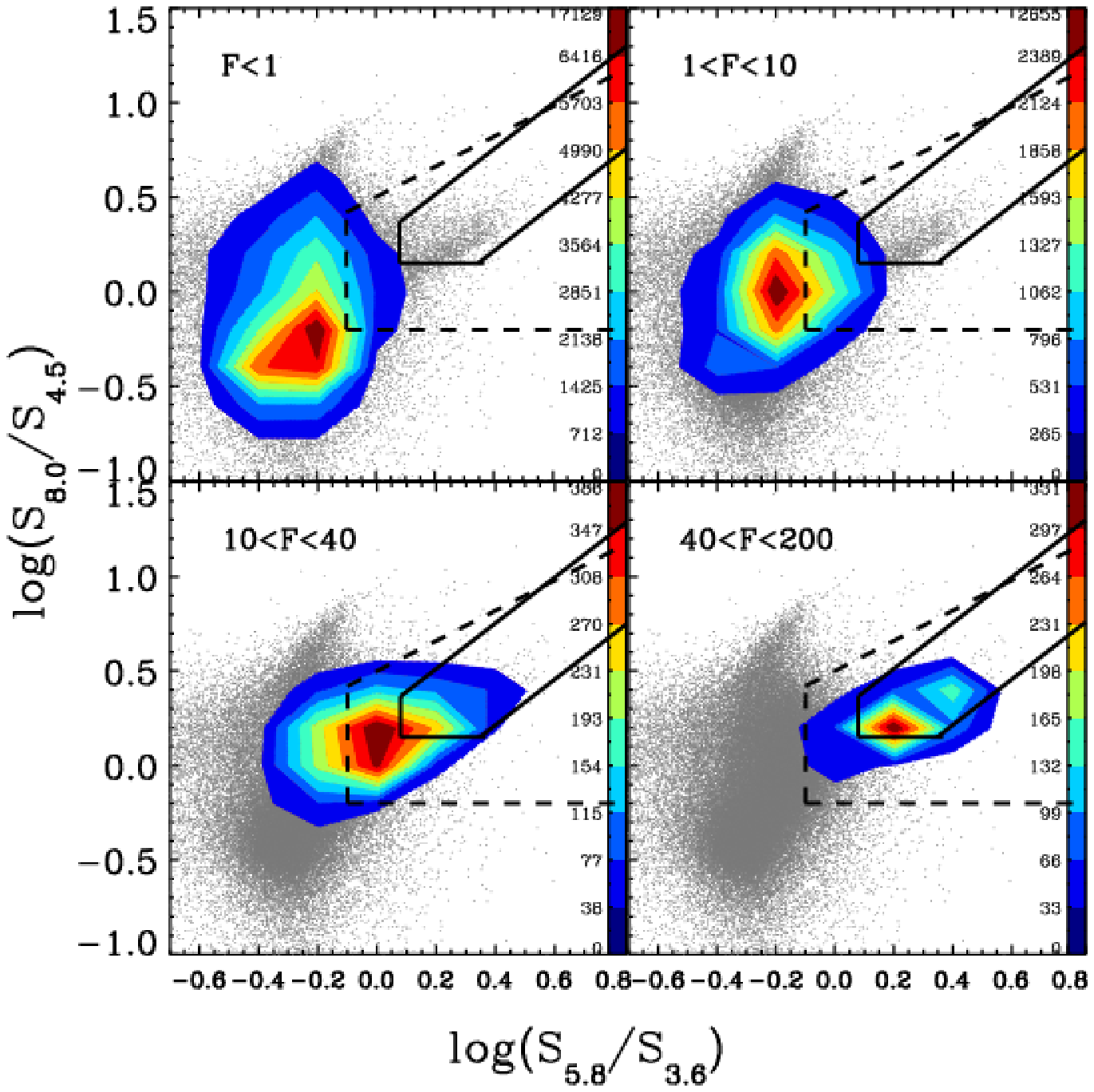}}
\caption{\scriptsize{ IRAC color-color diagram with contours showing the distribution of sources in increasing bins of $F$-ratio.  Black lines illustrate mid-IR selection criteria, as decsribed in Figure~\ref{fig:stern_wedge}. The grey cloud shows all non-stellar sources with $I<21$. }}
\label{fig:stern_wedge_fratio}
\end{figure*}

Figure~\ref{fig:stern_wedge} shows the IRAC colors for 77,277 non-stellar, $I<21$ sources in our sample, as compared to 1053 X-ray sources and 3114 $F>10$ sources.  Of the 1053 X-ray sources, there are 413 point-like ($S\geq0.7$) and 640 extended ($S<0.7$) sources.  The left panel shows the IRAC AGN selection criteria of \citet{Stern2005}, while the right panel shows the AGN selection criteria of \citet{Lacy2004} and \citet{Donley2012}.  The \citet{Stern2005} and \citet{Lacy2004} criteria are broadly similar, with the \citet{Stern2005} wedge including 62\% of all the X-ray AGN and the \citet{Lacy2004} wedge including 70\%.  The X-ray AGN that fall outside the \citet{Stern2005} and \citet{Lacy2004} wedge are mostly extended sources that have a significant luminosity contribution from their underlying host galaxies.   The X-ray sources that are AGN-dominated (i.e., optical point sources) are nearly all identified as mid-IR AGN, with $\sim$95\% falling within the \citet{Stern2005} and \citet{Lacy2004} wedges.  This is consistent with the results of \citet{Gorjian2008}, \citet{Cardamone2008}, and \citet{Mendez2013}, who find that mid-IR selection can miss a large fraction of X-ray identified AGN, especially moderate-luminosity or low accretion rate X-ray AGN.  The \citet{Donley2012} color criteria for AGN selection is more strict, and includes only 35\% of the overall X-ray AGN sample and 70\% of the point source X-ray AGN sample shown in Figure~\ref{fig:stern_wedge}.  The fractions of $F>10$ sources included in the \citet{Stern2005}, \citet{Lacy2004}, and \citet{Donley2012} AGN selection criteria are 57\%, 75\%, and 32\%, respectively.  The $F>10$ sample generally spans a wider range of IRAC colors than the X-ray sources.

To better understand how $F$-ratio is correlated with mid-IR color, we examine the IRAC color distribution of four different $F$-ratio samples, as shown in Figure~\ref{fig:stern_wedge_fratio}.   Sources with low $F$-ratios ($F<1$) are concentrated near $[3.6]-[4.5]\sim0$ and $[5.8]-[8.0]\sim0$ and form a tail out to redder $[5.8]-[8.0]$ colors.  As the $F$-ratio increases, the distribution moves towards redder colors in both $[3.6]-[4.5]$ and $[5.8]-[8.0]$, until at the highest $F$-ratio bin ($F>40$), the sources are almost exclusively located within the \citet{Stern2005} wedge.  

Figure~\ref{fig:stern_wedge_fratio} shows that the $F$-ratios are qualitatively consistent with the mid-IR color selection of AGN, in the sense that low $F$-ratios have colors consistent with quiescent galaxies, whereas high $F$-ratios have colors consistent with mid-IR AGN.  However, there is also an ``intermediate'' $F$-ratio population, many of which have bluer [3.6]-[4.5] colors.  A similar trend is seen in the right panel of Figure~\ref{fig:stern_wedge_fratio}, where IRAC colors become redder with increasing $F$-ratio, and the $F>40$ sample lying almost exclusively within the \citet{Lacy2004} wedge.  

\begin{figure}[!t] 
\centering
\includegraphics[scale=0.5]{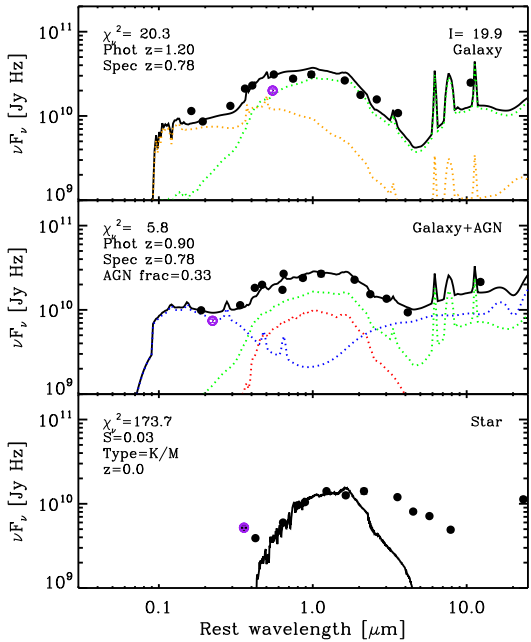}
\caption{\scriptsize{SED of an $F=10$ source ($\alpha=216.26442$ deg, $\delta=33.62389$ deg) with IRAC colors that place it just outside the \citet{Stern2005} wedge.  It is significantly better fit by a galaxy+AGN model (middle panel) than by a galaxy-only model (top) and its AGES spectrum contains a broad MgII emission line, confirming the presence of an AGN. }} 
\label{fig:SED_composite}
\end{figure}

\clearpage
Figure~\ref{fig:SED_composite} shows an example of an $F=10$ SED for a source with IRAC colors that lie just blueward of the \citet{Stern2005} and \citet{Lacy2004} wedges. The data are clearly better fit when an AGN component is included in the SED model.  The AGN component in this case contributes 33\% of the bolometric luminosity, while the rest of the luminosity is attributed to the star-forming and starburst galaxy SED models.  This is an example of an AGN with a host component that is strong enough that the mid-IR colors alone could not be used to identify this source as an AGN.  There is however, an AGES spectrum of this source, and the presence of a broad \ion{Mg}{2} $\lambda$2800 emission line confirms it as an AGN.  While this source does have an X-ray counterpart, it is not included in our X-ray sample because it does not meet the $\geq4$ counts criterion.

There are 86 sources with AGES spectra that have SED fits with $10<F<15$, and IRAC colors that are $\sim$0.2 mag bluer than the bottom edge of the \citet{Stern2005} wedge.  Of these 86 sources, only four were best-fit with a quasar template spectrum during the redshift cross-correlation procedure.  However, many composite AGN/star-forming galaxies would not be best fit by the quasar template because they lack strong, broad emission lines.  Using emission line ratio diagnostics, we can gain insight into the nature of the intermediate $F$-ratio SEDs that lie just outside the mid-IR AGN wedges.  
                  
\begin{figure*}[!t] 
\centering
\subfigure{\includegraphics[width=3in]{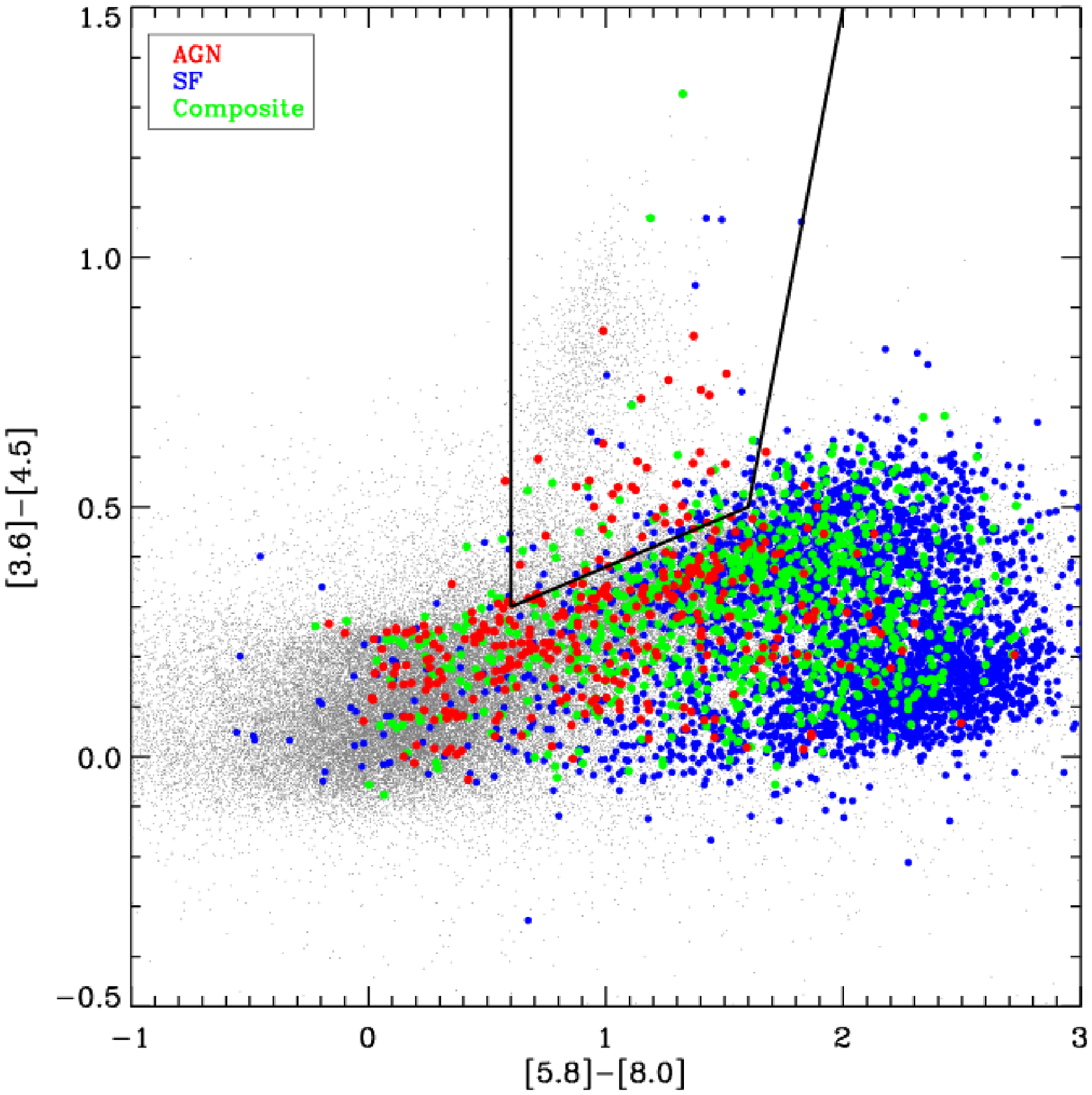}}
\subfigure{\includegraphics[width=3in]{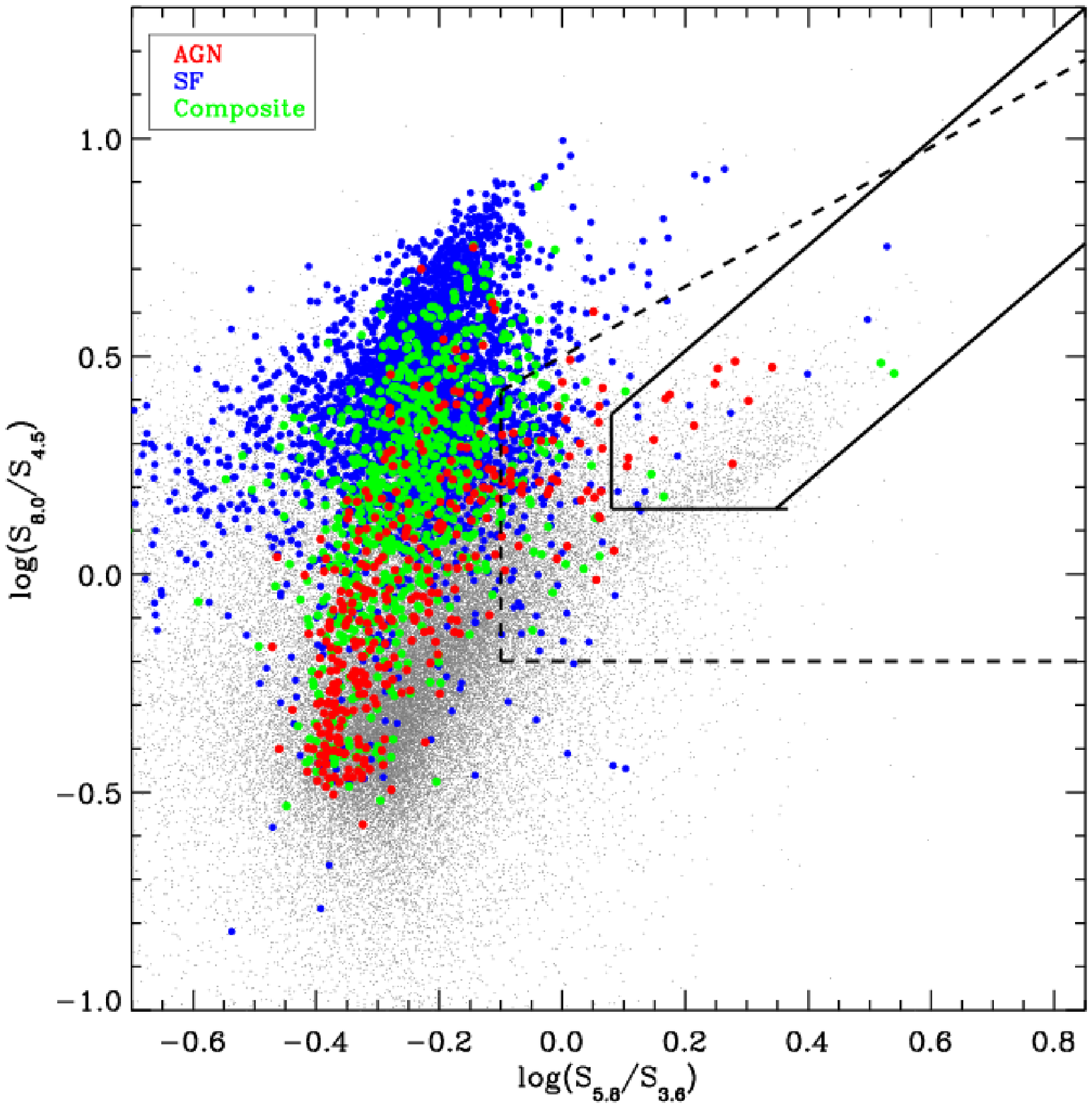}}
\caption{\scriptsize{IRAC color-color diagrams with sources classified as star-forming galaxies, AGN, or composite galaxies based on BPT diagnostic from emission line ratios, overplotted in blue, red, and green symbols, respectively.  Black lines illustrate mid-IR selection criteria, as described in Figure~\ref{fig:stern_wedge}.}}
\label{fig:stern_wedge_BPT} 
\end{figure*}   

Figure~\ref{fig:stern_wedge_BPT} again shows the IRAC mid-IR colors for non-stellar $I<21$ sources, with the various AGN wedges indicated in solid and dashed lines.  Objects that have been classified as either star-forming galaxies, AGN, or composite sources based on their Baldwin-Phillips-Terlevich \citep[BPT;][]{Baldwin1981} emission line ratios \citep{Moustakas2011}  are shown with the larger, colored symbols.   Note that the BPT diagnostic can only be used for low redshift $z_{s}\lesssim0.35$ narrow-line sources because the H$\alpha$ emission line at $\lambda6563$\AA\ is redshifted out of the AGES optical spectrum at $z\sim0.4$.  There are 3694, 920, and 420 star-forming galaxies, composite galaxies, and AGN shown in Figure~\ref{fig:stern_wedge_BPT}.  Among this sample of narrow emission line $z\lesssim0.35$ AGES galaxies and AGN, we  see that the star-forming galaxies and the AGN tend to occupy different, though overlapping, regions of IRAC color space, with the composite sources being clustered in between the two populations.  The BPT AGN and composite sources have a color distribution roughly parallel to the blue edge of the \citet{Stern2005} wedge.  This is similar to the color distribution of the $F>10$ sources that extend beyond the \citet{Stern2005} wedge, shown in Figure~\ref{fig:stern_wedge_fratio} and Figure~\ref{fig:stern_wedge_BPT}, strongly suggesting that the SED fits are identifying the narrow-line AGN and composite population.  Also note that the mid-IR colors of the narrow line AGN/composite sources are different from that of the X-ray sources.

\begin{figure*}[!tph] 
\centering
\includegraphics[scale=1]{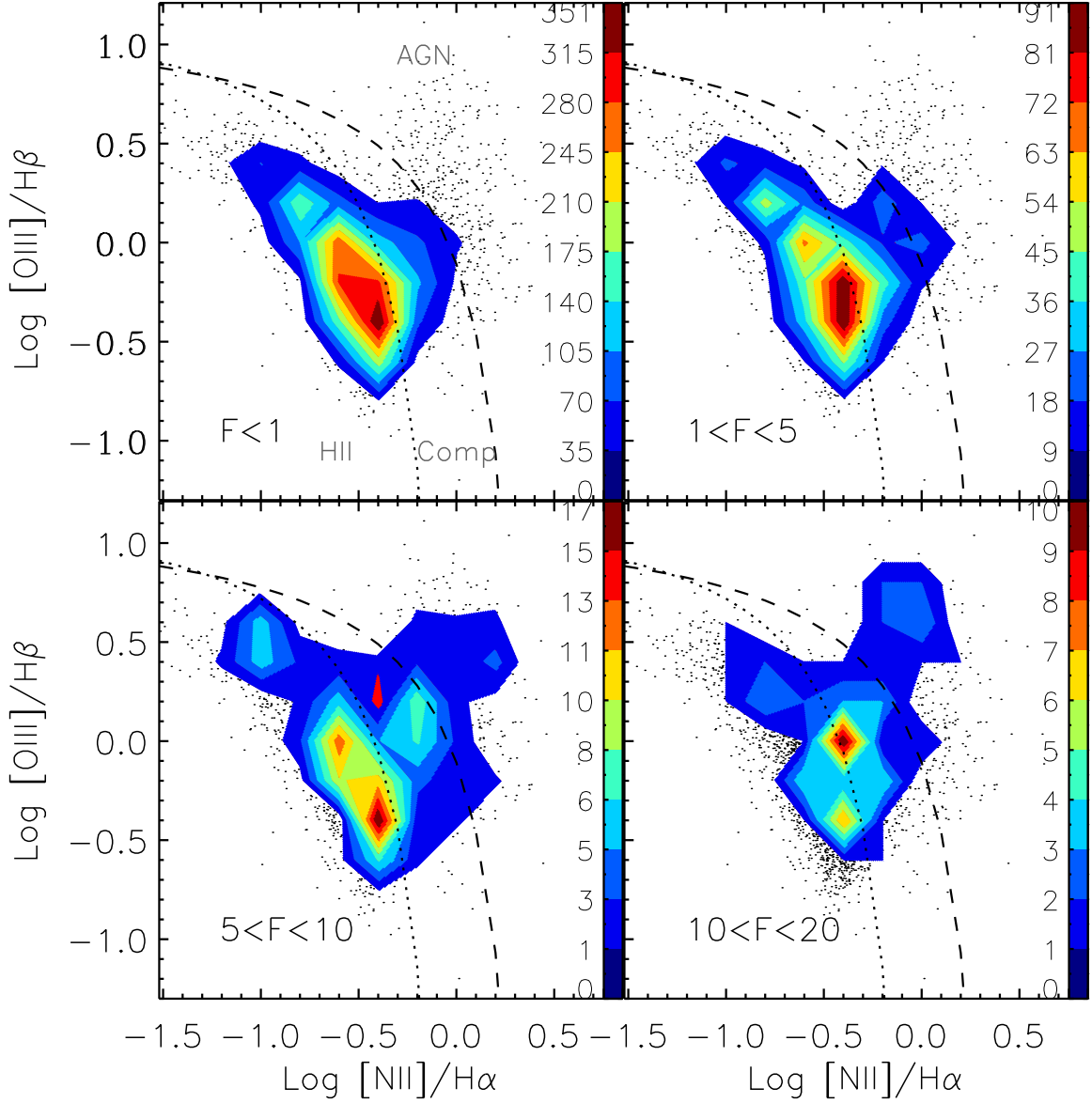}
\caption{\scriptsize{BPT diagram showing the emission line ratios of [OIII]/H$\beta$ and [NII]/H$\alpha$ of 4,030 sources.  Each panel shows the density contours of sources in different $F$-ratio bins.  The dotted and dashed lines demarcate regions dominated by star-forming galaxies (HII), composite sources, and narrow-line AGN.  As the $F$-ratio increases, there is a general shift towards the composite and narrow-lined AGN regions of the diagram. }}
\label{fig:BPT}
\end{figure*}

We examined the emission line ratios of 4,030 sources separated into bins of $F$-ratio in order to examine how the SED fitting technique compares to the BPT classification of narrow-line AGN, composite sources, and star-forming galaxies.  Figure~\ref{fig:BPT} shows the line ratio distribution of sources in $F$-ratio bins. The dotted and dashed lines demarcate regions on the BPT diagram where the emission line ratios can be explained by either \ion{H}{2} regions (i.e., star-forming galaxies) or AGN \citep{Kewley2006}.  Composite sources lie in the area between the two lines.  Figure~\ref{fig:BPT} does not show sources with $F>20$ because we want to focus on the ``ambiguous'' cases.  

Table~\ref{table:BPT} reports the distribution of these 4,030 sources in $F$-ratio and BPT classification.  For $F<1$, there are 2,973 sources and most of these (76\%) are classified as \ion{H}{2} regions and few (6\%) as AGN.  The AGN fraction then steadily increases as the $F$-ratio increases.  However, even for the $10<F<20$ bin, only 44\% of sources are classified as either a composite source or AGN.  This is an underestimate of the true AGN fraction because the BPT sample explicitly excludes higher redshift ($z\gtrsim0.35$), broad-lined AGN (Gaussian $\sigma>$ 500 km s$^{-1}$).

\begin{table*}[!t]
\caption{ \normalsize{BPT Classifications}}
\vspace{2mm}
\begin{center}
\begin{tabular}{c c c c c}
\hline\hline
\ & $F<1$ (2973) & $1<F<5$ (828) & $5<F<10$ (158) & $10<F<20$ (71) \\
\hline
HII       & 76\% & 67\% & 59\% & 56\% \\ 
Composite & 18\% & 20\% & 22\% & 18\% \\ 
AGN       & 6\%  & 13\% & 19\% & 26\% \\ 
\hline 
\end{tabular} 
\end{center}
\vspace{1mm}
\scriptsize Note. -- The entries are the percentages of sources in $F$-ratio bins with HII region, composite, or AGN emission line ratio classifications.  The total number of sources in each $F$-ratio bin is shown in parenthesis.
\label{table:BPT}
\end{table*}

\subsubsection{SDSS colors}

\begin{figure*}[!thp]
\centering
\includegraphics[scale=0.9]{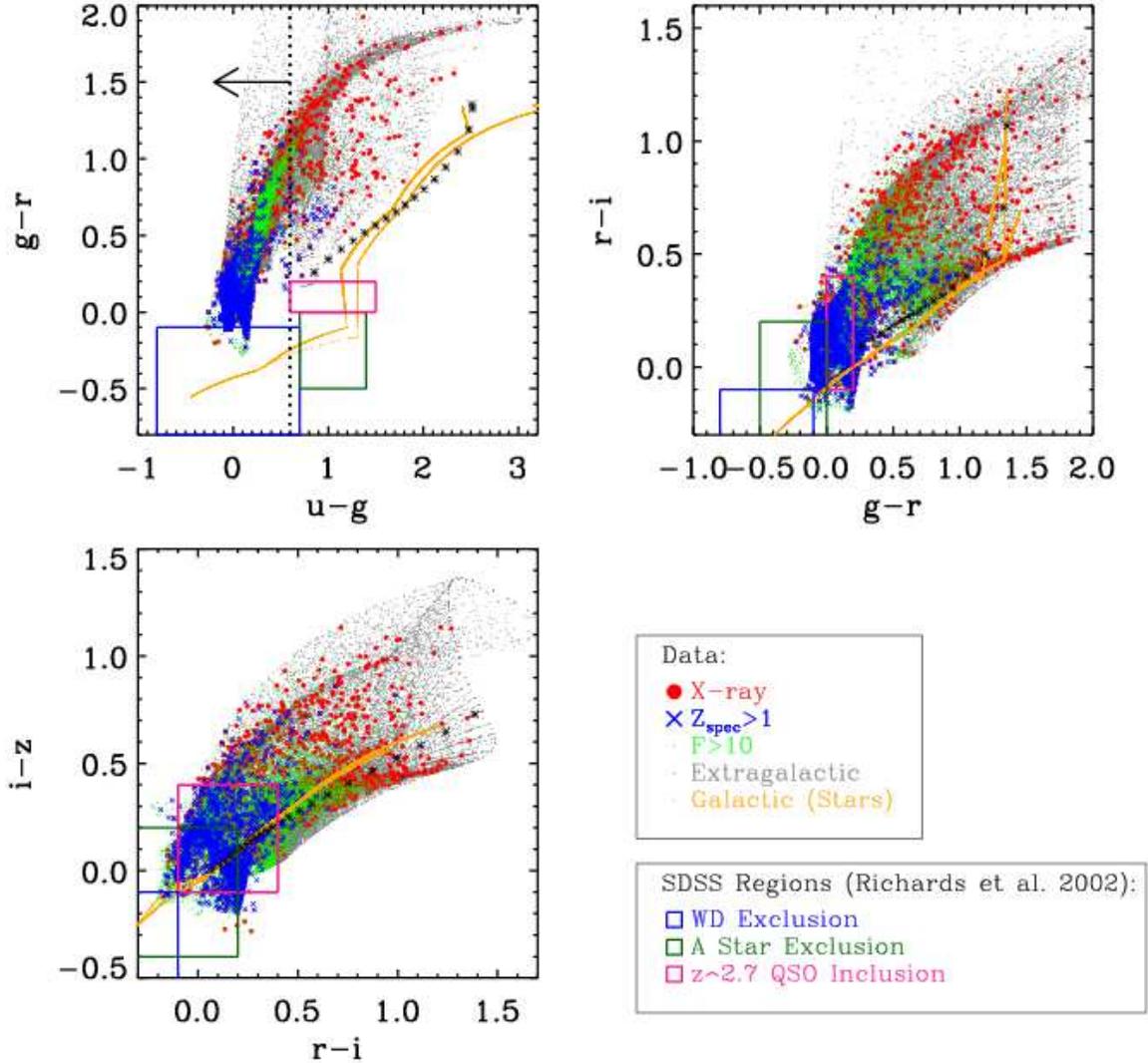}
\caption{\scriptsize{Synthesized SDSS colors for our sample of 383,604 extragalactic (gray dots) and 47,434 Galactic (orange dots) sources.  Because we synthesize the SDSS colors from the template fits, the orange stellar source sequences trace out two curves formed by the giant and dwarf templates.  The X-ray, $z_{s}>1$, and $F>10$ AGN samples are shown in red, blue, and green, respectively.  The SDSS exclusion and inclusion zones adopted from \citet{Richards2002} are shown, with the white dwarf  and A star exclusion zones shown as blue and dark green rectangles, and the  $z\sim2.7$ quasar inclusion zone is shown with the magenta rectangle.  Black asterisks show the SDSS stellar locus \citep{Richards2002}.  In the $g-r$, $u-g$ color space, the black dotted vertical line shows the equivalent UVX color criterion.  Sources that avoid stellar exclusion zones and are blueward of $u-g=0.7$ are considered quasar candidates.  Most of the AGN and AGN candidates from our sample meet the UVX color criterion.}}
\label{fig:sdss_colors}
\end{figure*}

In this section we examine the distribution of the sources in their optical colors.  Since we lack deep photometric data in the SDSS filters, we produced synthetic SDSS colors from the best fit SED models and the SDSS filter response curves.  Figure~\ref{fig:sdss_colors} shows the synthetic SDSS colors of the 383,604 extragalactic sources and 47,434 stars, where the Galactic sources are defined by \chisqred(Galaxy+AGN)$>$\chisqred(Star).  The X-ray, $z_{s}>1$, and $F>10$ samples are overplotted in red, blue, and pink, respectively.  The stars and brown dwarfs are shown in green.   

Quasar selection in SDSS is based on examining the colors of $i\lesssim20$ point sources and excluding known stellar regions of color space \citep{Richards2002}.  The blue and cyan boxes show the white dwarf and A star exclusion regions, and the green box indicates an inclusion region for $z\sim2.7$ quasars which also have some optical colors similar to A stars.  In the upper left panel of Figure~\ref{fig:sdss_colors} there is also a quasar inclusion region for sources bluer than $u-g=0.6$, but outside the white dwarf exclusion box.  This color cut is roughly equivalent to earlier UV excess (UVX) methods \citep[e.g.,][]{Boyle1990}. 

The SDSS stellar locus (black asterisks) is very similar to the synthesized SDSS colors of our stellar sample (green curves), confirming that the extragalactic/Galactic separation  based on \chisqred\ of the SED models is generally a success.  The synthesized SDSS colors of the stellar templates (green) form two distinct curves for dwarfs and giants, respectively, with each curve formed by the sequence of spectral types.

\begin{figure*}[!t] 
\centering
\subfigure{\includegraphics[width=3in]{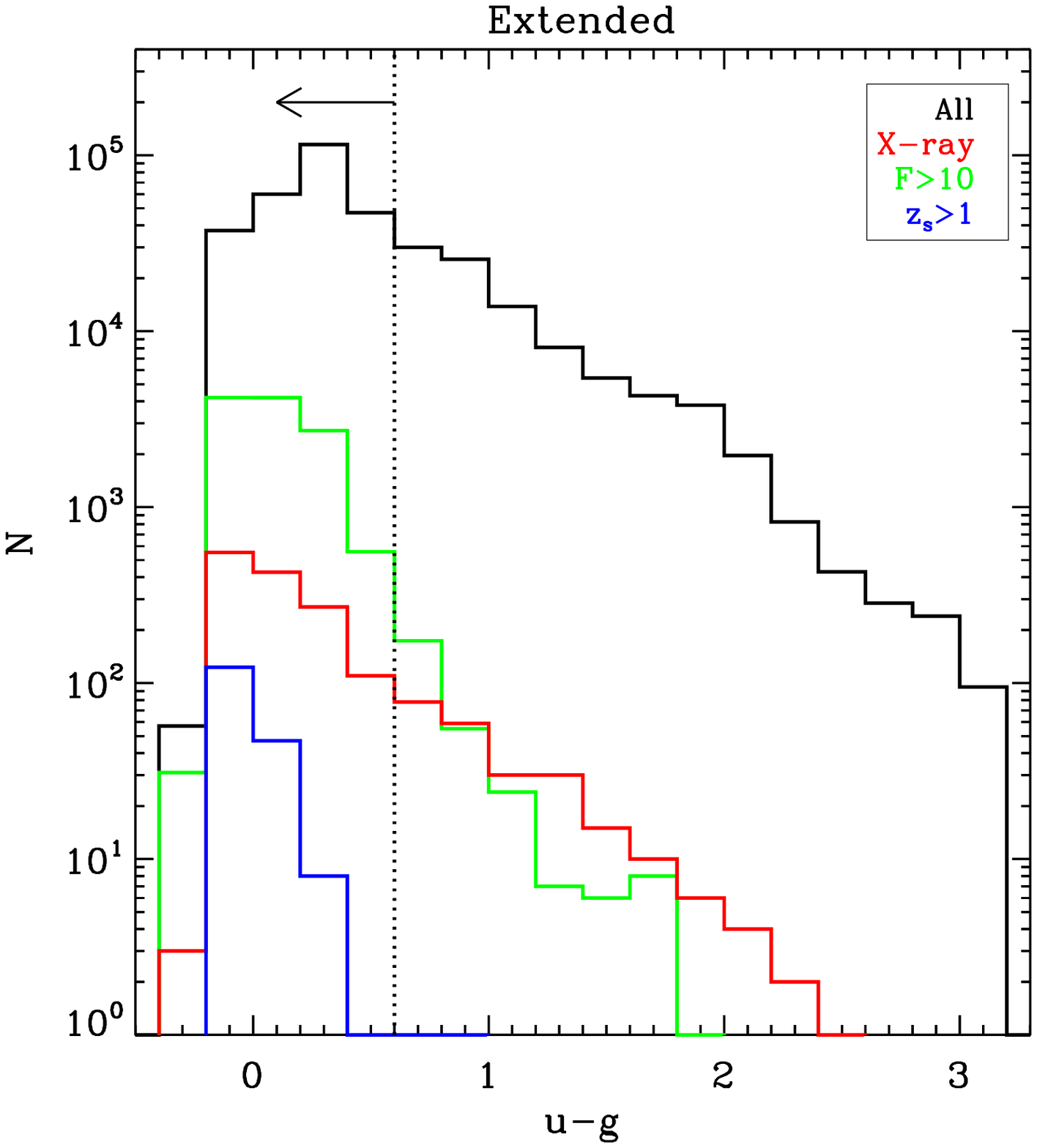}}
\subfigure{\includegraphics[width=3in]{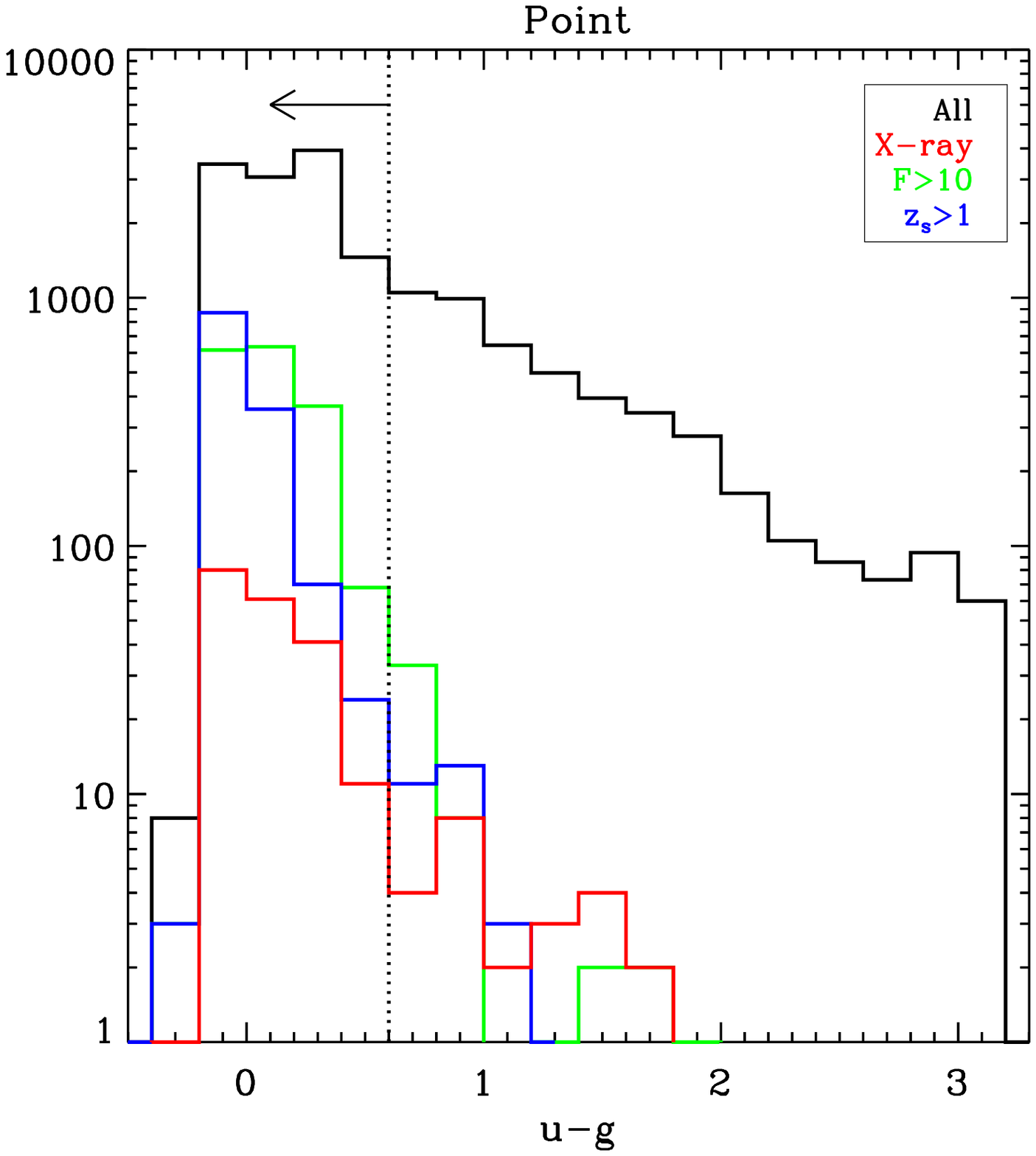}}
\caption{\scriptsize{Distribution of $u-g$ color among extended (left) and point-like (right) non-stellar sources.  The X-ray, $F>10$, and $z_{s}>1$ AGN candidate samples are shown in red, green, and blue histograms, respectively.  The vertical dotted line at $u-g=0.6$ highlights the SDSS UVX color criterion for AGN.}}
\label{fig:sdss_hist} 
\end{figure*}

In the upper right and lower panels of Figure~\ref{fig:sdss_colors}, the $z_{s}>1$ sample of AGN are roughly centered in the $z\sim2.7$ quasar inclusion region from \citet{Richards2002}, except in the $g-r$, $u-g$ color space where the $z_{s}>1$ and X-ray AGN do not coincide with the $z\sim2.7$ inclusion box.  Figure~\ref{fig:sdss_hist} shows the $u-g$ color distribution among the extended and point-like sources for our non-stellar sample, including the $z_{s}>1$, $F>10$, and X-ray sources.  The point-like AGN candidate samples show a relatively narrow $u-g$ color distribution in comparison to the extended sources, which is not surprising because point-like AGN have colors that are less contaminated by their host galaxies.  Among the extended sources, it is clear that both the X-ray and $F>10$ samples have $u-g$ colors that extend significantly beyond the $u-g<0.6$ UVX color criterion, demonstrating the potential of the $F$-ratio method to find a diverse sample of AGN.

\subsection{AGN Surface Density}
\label{sec:surface_density}

\begin{figure*}[!tp]   
\includegraphics[scale=0.8]{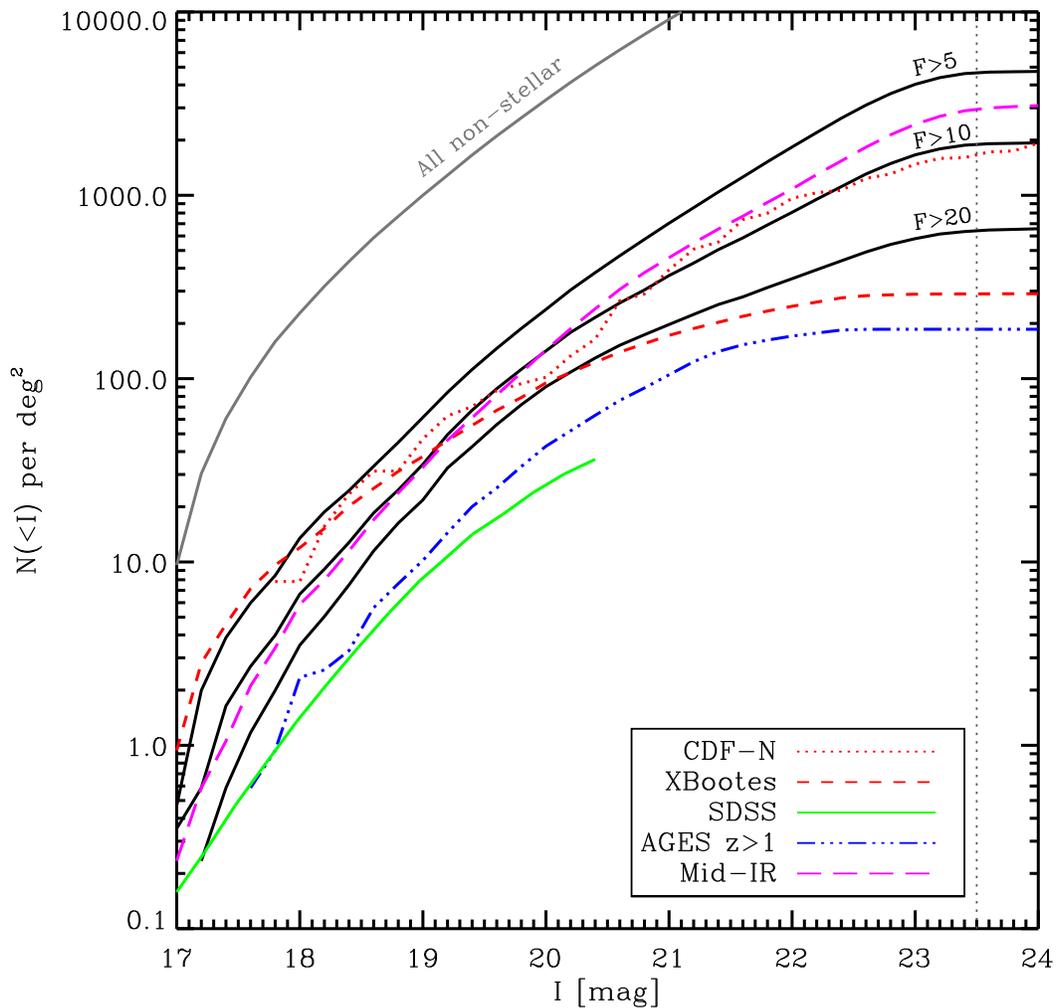}
\caption{\scriptsize{Integrated surface density of AGN candidates with $I>17$ (number per deg$^{2}$) as a function of $I$ magnitude.  The three curves in solid, dotted, and dashed black lines show the surface density of AGN candidates from the $F>5$, $F>10$, and $F>20$ samples.  For comparison, we show the surface densities of AGN candidates selected from X-ray, optical, and mid-IR surveys in red, green, and magenta, respectively.  The blue AGES sample is a spectroscopic sample of $z_{s}>1$ AGN. The vertical dotted line at $I=23.5$ shows the optical magnitude limit imposed on our sample.  The $F$-ratio selection of AGN yields higher AGN surface densities than the optically selected samples. }}
\label{fig:surface}
\end{figure*}

We can also examine the success of SED fitting as an AGN selection method by comparing the surface density of AGN candidates found from the $F$-test, optical, mid-IR, and X-ray selection methods. Figure~\ref{fig:surface} shows the integrated number of non-stellar sources in total and with $F>5$, $F>10$, and $F>20$ along with AGN surface densities from optical and X-ray surveys.  We also show the number of SDWFS sources that satisfy the \citet{Stern2005} mid-IR selection criteria.  As is typical of AGN distributions, the number rises steeply and then flattens at $I\sim20$. The further flattening at fainter magnitudes is due to completeness and the effects of steadily increasing photometric errors on $F$.

The surface density of point-like, optically selected quasars from SDSS and the 2dF/6dF QSO Redshift Survey (2QZ/6QZ) adopted from \citet{Richards2006} is shown by a solid green curve in Figure~\ref{fig:surface}.  These quasars were selected based on their optical (SDSS) colors and are also required to be point sources in the magnitude range of $18.0<g<21.85$.   We also show the number of $z_{s}>1$ AGES sources, which are dominated by point-like, mid-IR selected AGN with $I<21$ \citep{Kochanek2012}.  While different in selection methods, these sources are largely broad line quasars similar to what is found in the SDSS or 2QZ surveys.

In Figure~\ref{fig:stern_wedge_fratio} we showed that the \citet{Stern2005} mid-IR selection region typically contained sources with $F\gtrsim10$, and in Figure~\ref{fig:surface} we see that the surface density of $F>10$ sources is very similar to that of sources which satisfy the \citet{Stern2005} mid-IR selection criteria.  Mid-IR selection is relatively impervious to dust extinction and is not limited to optically point-like sources, so it is not surprising that it lies well above the SDSS/2QZ sample densities.  At faint magnitudes, the mid-IR sample is limited by the need for 5.8\micron\ and 8.0\micron\ detections, and the method fails as the AGN becomes similar in luminosity to its host, so it is still an incomplete inventory of AGN.  At least in AGES, there was very little contamination of the mid-IR sample by non-AGN, although this is likely a greater problem at $I>21$ where there begin to be galaxies that can match the $[3.6]-[4.5]$ criteria and the $[5.8]-[8.0]$ color is becoming noisier.  

Finally, we can compare to the shallow X\Bootes\ and deep \emph{Chandra} Deep Field-North \citep[CDF-N;][]{Brandt2001} X-ray selected samples.  These should identify all AGN other than the highly obscured, Compton thick population. For the CDF-N sample, we exclude sources with  low X-ray to optical flux ratios (i.e. $\log(f_{X}/f_{R})<-2$) because the X-ray emission of these sources is likely of stellar origin, and any objects that have been spectroscopically identified as stars.  Due to its small area, the CDF-N has few bright AGN, but for $18\lesssim I\lesssim20$, the X\Bootes\ and CDF-N surface densities are comparable.  The shallow X\Bootes\ survey is increasingly incomplete for $I>20$.  For $I\lesssim 19$, the density of the $F>5$ sample is similar to that of X-ray sources but then rises to be a factor of $\sim2$ higher at fainter magnitudes.  The $F>10$ sample is lower than the X-ray samples at bright magnitudes but has comparable number counts at the faint end. 

These comparisons of surface densities are consistent with the results in \S\ref{sec:compare}.  The $F\gtrsim 20$ sample will yield large numbers of luminous, broad-line AGN similar to those found in optical surveys.  The sources with $F\gtrsim 5-10$ will be similar to many of those found in X-ray surveys but with an increasing false positive rate for lower values of $F$.  The $10\lesssim F\lesssim20$ sources appear to track the composite population found by emission line diagnostics.  We find a total of 16,266 sources with $F>10$ at $I<23.5$, which yields a surface density of 1904 AGN deg$^{-2}$.  

\section{Summary and Future Work} 
\label{sec:future}

Using photometric data that ranges from the far-UV to the mid-IR, we fit galaxy, AGN, stellar, 
and brown dwarf SED models to 431,038 sources in the \Bootes\ NDWFS field.  The photometric
separation of stars and galaxies compares well with morphological separation and star and
galaxy surface densities at all magnitudes.  Comparing to the over 20,000 available
spectroscopic redshifts, we find photometric redshift dispersions of $\sigma/(1+z)$=0.040 
and $\sigma/(1+z)=0.169$ for the galaxy and AGN samples, respectively, after clipping the 
worst 5\% of sources.  In practice, the AGES survey \citep{Kochanek2012} obtained 
spectroscopic redshifts for a large fraction of the most problematic sources, the
luminous broad line AGN, so the photometric redshift dispersion of the sources without
spectra should be lower.  We estimated the likelihood of an AGN component based on F-test
comparisons of the fits.  The results for all the sources are reported in Table~\ref{table:results}.

When we examine the distribution of F-ratios as a function of morphology and host
luminosity fraction, we see the expected trends that high F-ratio sources tend to
be more point-like and have smaller host galaxy contributions to their SEDs.  X-ray
sources show a broader distribution of F-ratios and host galaxy contributions.
We also examined the distribution of sources relative to the \citet{Stern2005}, \citet{Lacy2004}
and \citet{Donley2012} mid-IR selection criteria.  As a comparison, 62\%, 70\% and 35\%
of the X\Bootes\ X-ray sources satisfy these criteria.  There is a clear trend of the
sources moving into these selection regions as $F$ increases, with 57\%, 75\% and
32\% of the sources with F-ratios $F>10$ for adding an AGN component falling in 
these mid-IR selection regions.  Sources with $F>40$ lie almost exclusively in these
regions.  The $F \sim 10$ sources have significant host contributions, and many
lie just bluewards of these selection regions in their $[3.6]-[4.5]$ colors. They
partially overlap the mid-IR color distribution of X-ray sources, but they extend
towards redder $[5.8]-[8.0]$ colors.  Their mid-IR color distribution is very 
similar to that of narrow line sources with ``composite'' line ratios indicative
of a mixture of star formation and AGN activity, which is also consistent with
their redder $[5.8]-[8.0]$ colors.  Similarly, we can use the F-ratio distribution
to examine how AGN extend outwards from the SDSS optical color selection regions  
as the host galaxy contribution becomes more important.

These results strongly suggest that a sample of $F>10$ sources can provide a
more complete inventory of AGN activity than any of the methods restricted
to limited wavelength regimes.  Like any method, it is not perfect. There are
clearly a minority of X-ray AGN with such low F-ratios that they cannot be 
identified based on their UV-IR SEDs with an acceptable false positive rate.
What needs to be calibrated at this point are the false positive rates as
a function of the apparent significance of the AGN contribution.  The
surface densities of $F>10$ sources as compared to all extragalactic sources
and other AGN samples suggests that the false positive rate should be
acceptably low, but this needs to be tested with spectroscopy.  For
$F\sim 5$, the false positive rate is likely unacceptably high.  The
problem for spectroscopy is that many redshift ranges will lack the 
emission line diagnostics needed to classify the nature of the source.

Most surveys of evolution separate the study of galaxies and AGN because
they have difficulties tracking the populations which are strong mixtures
of both.  This large scale decomposition of the SEDs into host and AGN 
components provides a means of studying the co-evolution of these 
populations with relatively reliable estimates of the two components
separately.  Even when the models have only photometric redshifts,
\citet{Assef2010} found that the estimates of the host galaxy luminosity
fraction were relatively robust.  In particular, it should be feasible
to examine the duty cycle of AGN activity in galaxies as a function of
redshift.  For example, if the false positive rates for AGN activity
can be calibrated as a function of $F$ using modest spectroscopic samples,
then the full survey sample can be used in studies of the evolution
of galaxies and AGN.

These very broad baseline SED models should be comparably ``stable'' to
surveys using large numbers of narrower filters over smaller wavelength
ranges \citep[e.g.,][]{Geach2008,Abramo2012}.  Instead of trying to better identify spectral breaks
or strong emission lines, the broad structure of galaxy SEDs with a peak
in the near-IR allows robust photometric redshifts and the structure of 
the UV and mid-IR tails of the distribution provides a robust probe of 
star formation and AGN activity.  This approach should be particularly
valuable for integrating wide area, multi-wavelength surveys in the
ultraviolet (\emph{GALEX}), optical (SDSS, DES\footnote{http://www.darkenergysurvey.org/}, Pan-STARRS\footnote{http://pan-starrs.ifa.hawaii.edu/public/}, LSST\footnote{http://www.lsst.org/}), near-IR (VISTA, \citet{Emerson2006}, \emph{EUCLID}\footnote{http://sci.esa.int/euclid/45403-mission-status/}, \emph{WFIRST}\footnote{http://wfirst.gsfc.nasa.gov/}) and mid-IR (\emph{WISE}, \citet{Wright2010}, \emph{Akari} \citep{Murakami2007}).

\section*{Acknowledgement}

 The authors would like to thank the \Bootes\ collaborations for contributing to the various \Bootes\ photometric catalogs.   The work of D.S. was carried out at Jet Propulsion Laboratory, California Institute of Technology, under a contract with NASA.  RJA was supported by Gemini-CONICYT grant number 32120009. 

\begin{landscape} 
\begin{deluxetable}{ccrrrrrrrrrrrrrrrrrrr}
\tabletypesize{\tiny}
\tablecolumns{21}
\tablewidth{0pt}
\tablecaption{\normalsize {Model results}}
\tablehead{
\colhead{}   \\
\colhead{RA}  & 
\colhead{Dec} & 
\colhead{$I$} & 
\colhead{$S$} & 
\colhead{z$_{\rm s}$} &   
\colhead{N}\vline   &   
\colhead{z$_{\rm p}$} &    
\colhead{\chisqred} & 
\colhead{L$_{\rm Ell}$} &  
\colhead{L$_{\rm Sbc}$} &  
\colhead{L$_{\rm Irr}$}\vline &  
\colhead{z$_{\rm p}$} &  
\colhead{\chisqred} & 
\colhead{L$_{\rm AGN}$} &  
\colhead{L$_{\rm Ell}$} &  
\colhead{L$_{\rm Sbc}$} &  
\colhead{L$_{\rm Irr}$} &  
\colhead{$F$} & 
\colhead{E(B-V)}\vline & 
\colhead{\chisqred} & 
\colhead{Stellar}   \\
\multicolumn{1}{c}{[deg]} & 
\multicolumn{1}{c}{[deg]} & 
\multicolumn{1}{c}{[mag]} & 
\multicolumn{3}{c}{}\vline & 
\multicolumn{2}{c}{} & 
\multicolumn{1}{c}{Galaxy } \hspace{-6mm} &
\multicolumn{1}{l}{Only} &
\multicolumn{1}{c}{}\vline &
\multicolumn{3}{c}{} & 
\multicolumn{1}{c}{Galaxy} \hspace{-9mm} &
\multicolumn{1}{c}{+AGN} & 
\multicolumn{1}{c}{} & 
\multicolumn{2}{c}{}\vline & 
\multicolumn{1}{c}{} & 
\multicolumn{1}{c}{Type} \\
}
\startdata
217.43651 &  34.12883 &  17.70 &   0.99 & \ldots &   13/1 &   0.47 &  36.08 &  $ 1.416$ &      Zero &      Zero &   0.47 &  45.10 &      Zero &  $ 1.416$ &      Zero &      Zero &   0.00 &   0.00 &   9.50 &        K/M \\ 
219.23635 &  34.12883 &  22.68 &   0.42 & \ldots &   14/2 &   0.60 &   0.98 &      Zero &  $-1.463$ &  $-0.148$ &   0.55 &   0.79 &  $-0.802$ &      Zero &  $-0.861$ &  $-0.366$ &   2.42 &   0.00 &   2.12 &        f/g \\ 
219.53921 &  34.12883 &  20.77 &   0.03 & \ldots &   12/3 &   0.25 &   1.52 &  $-1.114$ &      Zero &  $-0.407$ &   0.25 &   1.86 &      Zero &  $-1.114$ &      Zero &  $-0.407$ &   0.00 &   0.00 &  14.10 &        f/g \\ 
218.86312 &  34.12883 &  22.27 &   0.02 & \ldots &   15/0 &   1.80 &   1.02 &      Zero &  $ 1.191$ &  $ 1.313$ &   1.50 &   1.21 &  $ 1.275$ &  $ 0.133$ &  $ 1.053$ &  $ 0.233$ &   0.15 &   0.08 &  28.95 &        f/g \\ 
219.12657 &  34.12884 &  20.00 &   0.03 & \ldots &   14/1 &   0.54 &   3.24 &  $ 0.598$ &  $ 0.474$ &      Zero &   0.53 &   3.84 &  $ 0.027$ &  $ 0.711$ &  $-0.617$ &  $-0.594$ &   0.14 &   0.40 &  28.61 &      2000K \\ 
218.70934 &  34.12884 &  22.97 &   0.37 & \ldots &   14/0 &   1.54 &   1.17 &  $ 0.043$ &  $ 0.259$ &  $ 1.042$ &   1.53 &   0.57 &  $ 1.074$ &  $ 0.652$ &      Zero &  $-0.954$ &   6.23 &   0.04 &   6.18 &        a/f \\ 
218.04767 &  34.12886 &  21.91 &   0.73 & \ldots &   12/3 &   0.24 &   0.89 &  $-0.918$ &      Zero &      Zero &   0.24 &   1.09 &      Zero &  $-0.918$ &      Zero &      Zero &   0.00 &   0.00 &   0.59 &        K/M \\ 
216.28490 &  34.12886 &  19.51 &   0.98 & \ldots &   14/0 &   0.39 &  25.40 &  $ 0.551$ &      Zero &      Zero &   0.39 &  32.66 &      Zero &  $ 0.551$ &      Zero &      Zero &   0.00 &   0.00 &   7.02 &        K/M \\ 
217.74110 &  34.12886 &  22.52 &   0.17 & \ldots &   12/2 &   1.93 &   2.06 &      Zero &      Zero &  $ 1.367$ &   2.29 &   1.55 &  $ 1.671$ &  $ 0.827$ &      Zero &      Zero &   2.68 &   0.05 &   3.96 &        f/g \\ 
219.01851 &  34.12886 &  21.68 &   0.16 & \ldots &   13/2 &   0.61 &   1.24 &      Zero &  $ 0.479$ &      Zero &   0.61 &   1.52 &      Zero &      Zero &  $ 0.479$ &      Zero &   0.00 &   0.00 &   5.65 &        K/M \\ 
216.70422 &  34.12887 &  21.90 &   0.03 & \ldots &   12/2 &   0.73 &   1.33 &      Zero &  $-0.789$ &  $ 0.319$ &   0.65 &   1.45 &  $-0.533$ &  $-0.802$ &      Zero &  $ 0.121$ &   0.60 &   0.00 &   4.71 &        f/g \\ 

\enddata
\label{table:results}
\tablecomments{RA/Dec are the source coordinate in decimal degrees, $I$ is the 6\farcs0 diameter aperture magnitude, $S$ is the SExtractor stellarity
index, $z_s$ is the AGES spectroscopic redshift if available and N is the number of bands with detections/limits used in the fits.  For
the Galaxy and Galaxy+AGN fits, $z_p$ is the photometric redshift, $\chi^2_\nu$ is the goodness of fit, L$_{\rm Ell}$, L$_{\rm Sbc}$,
L$_{\rm Irr}$ and L$_{\rm AGN}$ are the luminosities of the template components in units of $\log(L/10^{10}L_\odot)$.  Luminosities are calculated
for the fit at the spectroscopic redshift if known, and an entry of ``Zero'' means that the best fit included no contribution
from that template.  The Galaxy+AGN section also include the reddening $E(B-V)$ applied to the AGN template and the $F$-ratio
value compared to the Galaxy-only fit.  The stellar fits include $\chi^2_\nu$ and the best fit template, where $K/M$ means the best fit lay between the K and M stellar templates and $2000$~K means the best fit was the $2000$~K brown dwarf template.  Uppercase letters refer to giant spectral templates and lowercase letters refer to dwarf spectral templates. If used in other contexts, the spectroscopic redshifts and photometry should be referenced to \citet{Kochanek2012} and \citet{Brown2007}, respectively.}

\end{deluxetable}
\clearpage
\end{landscape}

\small

\bibliography{AGNselect}{}
\bibliographystyle{plain}

\end{document}